\documentclass{jfm}

\usepackage{graphicx}
\usepackage{newtxtext}
\usepackage{newtxmath}
\usepackage{natbib}
\usepackage{hyperref}
\hypersetup{
    colorlinks = true,
    urlcolor   = blue,
    citecolor  = black,
}

\newcommand{\RomanNumeralCaps}[1]
\linenumbers

\title{Parametric reduced-order modeling and mode sensitivity of actuated cylinder flow from a matrix manifold perspective}

\author{Shintaro Sato\aff{1}
  \corresp{\email{shintaro.sato.c3@tohoku.ac.jp}}
 \and Oliver T. Schmidt\aff{2}}

\affiliation{\aff{1}Department of Aerospace Engineering, Tohoku University, Aramaki-aza-Aoba 6-6-01, Aoba-ku, Sendai 980-8579, Japan
\aff{2}Department of Mechanical and Aerospace Engineering, University of California San Diego, CA 92093, USA}

\begin{document}
\maketitle

\begin{abstract}
  We present a framework for parametric proper orthogonal decomposition (POD)-Galerkin reduced-order modeling (ROM) of fluid flows that accommodates variations in flow parameters and control inputs.
  As an initial step, to explore how the locally optimal POD modes vary with parameter changes, we demonstrate a sensitivity analysis of POD modes and their spanned subspace, respectively rooted in Stiefel and Grassmann manifolds.
  The sensitivity analysis, by defining distance between POD modes for different parameters, is applied to the flow around a rotating cylinder with varying Reynolds numbers and rotation rates.
  The sensitivity of the subspace spanned by POD modes to parameter changes is represented by a tangent vector on the Grassmann manifold.
  For the cylinder case, the inverse of the subspace sensitivity on the Grassmann manifold is proportional to the Roshko number, highlighting the connection between geometric properties and flow physics.
  Furthermore, the Reynolds number at which the subspace sensitivity approaches infinity corresponds to the lower bound at which the characteristic frequency of the K\'arm\'an vortex street exists (Noack \& Eckelmann, {\it J. Fluid Mech.}, 1994).
  From the Stiefel manifold perspective, sensitivity modes are derived to represent the flow field sensitivity, comprising the sensitivities of the POD modes and expansion coefficients.
  The temporal evolution of the flow field sensitivity is represented by superposing the sensitivity modes.
  Lastly, we devise a parametric POD-Galerkin ROM based on subspace interpolation on the Grassmann manifold.
  The reconstruction error of the ROM is intimately linked to the subspace-estimation error, which is in turn closely related to subspace sensitivity.
\end{abstract}



\section{Introduction}
Gaining key insights into fluid flow behavior enhances our understanding of fluid dynamics
and further develops fluid mechanics for the accurate prediction and active control of fluid flows.
Data-driven approaches for capturing the essential features of fluid flows have been intensively developed
owing to significant improvements in experimental techniques and numerical simulations \citep{Rowley2017,Taira2017}.
Proper orthogonal decomposition (POD) is one of the most common approaches for extracting flow structures that characterize the flow field of interest \citep{Lumley}.
POD analysis seeks the basis, referred to as POD modes, which optimally
represents the fluctuation component of the time-series dataset of the flow field obtained through experiments or numerical simulations.
Dynamic mode decomposition (DMD) is a common modal-decomposition method
that estimates a linear operator approximating the dynamics of interest from time-series snapshot data \citep{PSchmid2010}.
Each DMD mode has a single frequency in terms of time and growth rate.
Both modal-analysis techniques have been extensively employed to understand, predict, and control fluid flows.
The spectral modal analysis explored in recent years, such as spectral POD \citep{Towne2018}, has successfully extracted coherent structures from turbulent flows.

Modal decomposition, particularly for POD, postulates that the time series of snapshot fluid flow data,
represented as a state vector in a high-dimensional Euclidean space, lies in an inherently low-dimensional subspace.
Hereafter, the discussion assumes the use of the POD for modal analysis.
As POD analysis identifies an optimal subspace, POD modes are ideally suited for constructing a reduced-order model (ROM) of the fluid-flow dynamics.
The Galerkin projection-based ROM, which consists of ordinary differential equations (ODEs) of the expansion coefficients with respect to time
is a well-known technique for constructing the ROM \citep{Noack2011}.
Because the ROM approximates the dynamics of the fluid flow in a low-dimensional subspace,
the computational cost is significantly lower than that required when solving the full-order model, which is typically the Navier--Stokes equation.
Consequently, the ROM has been employed to predict the future state of the flow field and actively control fluid flows \citep{Taylor2004}.

However, a major challenge is that a typical ROM fails to provide an appropriate solution for flow parameters
(e.g., Reynolds number, Mach number, and object shapes, such as a cylinder and airfoils) that is different from that of the parameters under which the POD analysis is performed.
This is because POD analysis obtains the optimal set of bases to represent time-series snapshot data under the specific flow conditions analyzed.
The optimal bases can vary with the flow conditions \citep{Sato2021}.
The parametric ROM, whose objective is to perform multiple simulations over a wide range of parameter settings with a low computational cost,
has been examined in numerous studies on applications such as optimal design, uncertainty quantification, and active control \citep{Benner2015}.
A straightforward approach for constructing a parametric ROM is to extract optimal modes from a mixed database that contains datasets with different parameters \citep{Ma2002}.
\cite{Galletti2004} reported that a flow field around a square cylinder was appropriately reproduced at Reynolds numbers not included in the dataset used for modal decomposition.
\cite{Nakamura2024} demonstrated a parametric ROM that predicts flow fields around various object geometries by performing POD in computational space instead of physical space.
Although these global modes can be used to construct parametric ROM,
the global modes are extracted for global optimization using datasets for all parameters, but are not locally optimal.
Generally, the dimensions of the subspace increase with the number of parameters to be included,
resulting in an increase in the computational cost of solving ODEs for the ROM.

To construct a parametric ROM with a low computational cost, the dimensions of the subspace should be kept low.
This motivated us to investigate and model how the locally optimal subspace, which is obtained by conventional POD analysis, varies with the parameters
because a parametric ROM can be constructed by estimating the locally optimal low-dimensional subspace as a function of these flow parameters.
This study focuses on a geometric methodology that describes the parameter dependence of POD modes and subspace spanned by the POD modes.
We assume that the subspace exhibits small variations for small parameter changes; thus, the subspace is a continuous function of the parameters.
To discuss this continuity, it is important to define and compute the distance between subspaces.

To this date, the subspaces extracted through the POD analysis were typically examined independently for each parameter.
By contrast, this study aims to analyze the relationships between subspaces corresponding to different parameters by defining the distance between them.
The set of all $r$-dimensional subspaces of an $n$-dimensional vector space admits a manifold structure ($r$ and $n$ are natural numbers such that $r<n$),
which is referred to as the Grassmann manifold \citep{Edelman1998}.
Consequently, this is a natural method to consider the distance, defined on the Grassmann manifold, between two subspaces spanned by POD modes with different parameters.

The objective of this study is to develop a geometric methodology for analyzing the relationships between the local characteristic structures of fluid flows (i.e., POD modes)
and the subspaces spanned by them,
which are extracted from the datasets obtained by numerical simulations of the Navier--Stokes equations for parametric variations in the boundary conditions.
The distance between two points on the Grassmann manifold enables a quantitative discussion of the similarity between the subspaces corresponding to two different parameters.
Moreover, the parameter dependence of the subspace can be represented by modeling the curves or curved surfaces on the Grassmann manifold,
enabling the construction of a parametric ROM derived from this representation.

The first part of this study focuses on investigating how the POD modes and the subspace spanned by them vary with flow parameters.
We discuss the parameter dependence of the subspace spanned by the POD modes in terms of the geometric characteristics of a curve or curved surface on the Grassmann manifold.
The sensitivity of the subspace is defined as the ratio of the subspace displacement along the curve to the parameter change.
We show that the subspace sensitivity is closely related to the change in the behavior of fluid flows with respect to the variation in the parameters.
In addition, a method for visualizing the distribution of subspaces over a wide range of parameters on the Grassmann manifold is explored.

Subsequently, we discuss the parameter dependence of the POD modes rather than the subspace.
Parametric analysis of the set of POD modes for each parameter is conducted on the Stiefel manifold,
which is the set of all $n$-by-$r$ orthonormal matrices \citep{Edelman1998}; that is, the set of POD modes for each parameter is an element of the Stiefel manifold.
Manifolds that have a natural representation of elements in the form of a matrix are referred to as matrix manifolds \citep{Absil}.
Matrix manifolds provide insights into the intrinsic geometric properties of subsets of matrices defined by specific constraints (e.g., orthogonality).
We examine the sensitivity of the set of POD modes with respect to parameter changes from the perspective of the Stiefel manifold.
\cite{Hay2009} derived the sensitivity of the POD modes by differentiating the POD modes using the equation used in the method of snapshots developed by \cite{Sirovich1987},
and demonstrated that considering the subspace spanned by the POD modes and POD-mode sensitivity improves the performance of the parametric ROM.
In this study, the sensitivity of the POD modes is defined as the tangent vector along a curve on the Stiefel manifold,
and the variation in the POD modes with the parameter changes is investigated.
We then introduce sensitivity modes based on the analysis of the POD modes on the Stiefel manifold to represent the sensitivity of the flow field
with respect to the parameter changes by superimposing the sensitivity modes.
The analysis of flow sensitivities enables the estimation of the flow field at nearby parameters of interest, such as the operating or design parameters of the fluid machinery,
as well as the estimation of the response of the flow field to input-parameter uncertainty \citep{Pelletier2008}.
The sensitivity modes of the flow field can be interpreted as modes that represent the changes in the flow field in response to small variations in the parameters.
We demonstrate that the sensitivity of the flow field can be characterized by analyzing the sensitivity of the POD modes.

The last part of this study examines the parametric ROM, which employs the subspace-interpolation technique on the Grassmann manifold.
Interpolation of the subspaces spanned by the POD modes at different parameters was employed to construct a parametric ROM.
\cite{Lieu2007} developed a parametric ROM framework by interpolating two subspaces based on the principal angles between them.
\cite{Amsallem2008} generalized this framework as the subspace interpolation method based on the Grassmann manifold
and applied the developed framework to an aeroelastic ROM using sets of POD modes obtained using the Euler equation 
for a full-order model of the fluid flow around an aircraft.
Recently, \cite{Pawar2020} developed a parametric ROM framework combining the subspace interpolation on the Grassmann manifold
and a long short-term memory neural-network architecture to predict unknown physics that was not included in the training dataset
through demonstrations of the Burgers and vorticity-transport equations.
\cite{Hess2023} employed the manifold-interpolation technique to interpolate the linear operator estimated by the DMD analysis,
and demonstrated that the Reyleigh-B\'enard cavity flows over a wide range of the Grashof numbers can be reconstructed using the developed parametric ROM framework.

This study explores the relationship between the geometric characteristics of the distribution of subspaces on the Grassmann manifold
and the reconstruction errors obtained by the parametric ROM using subspace interpolation over a wide range of flow parameters.
We demonstrate that the reconstruction error of the flow field obtained by the parametric ROM is closely related to the estimation error of the subspace,
which is associated with the sensitivity of the subspace to parameter variations.
This finding is critical for constructing a parametric ROM using fewer samples.

The remainder of this paper is organized as follows.
Section~\ref{Sec:Method} describes the computational methods for the mode sensitivity analysis and ROM on the matrix manifolds.
In \S\ref{Sec:SensAnalysis}, we discuss the sensitivity analysis of the subspace, POD modes, and flow field based on the geometric structure of the matrix manifolds
using the flow field data around a rotating cylinder as a demonstration.
The parametric ROM is performed to discuss the relationship between the reconstruction error and geometric characteristics of a curve or curved surface on the Grassmann manifold in \S\ref{Sec:pROM}.
Finally, Section~\ref{Sec:Conclusions} presents the conclusions of this study.

\section{Methodology}\label{Sec:Method}
\subsection{Overview of mode sensitivity analysis on matrix manifolds}
Matrix manifolds play an important role in various applications, including parametric-model reduction for nonlinear dynamics, which is not limited to fluid dynamics \citep{Son2013,Liu2022},
Matrix-completion problems \citep{Boumal2015}, and computer vision \citep{Lui2012}.
Here, we discuss the basic idea of the parametric modal analysis for investigating mode sensitivity
and constructing parametric ROM on matrix manifolds before discussing specific manifolds.
\begin{figure}
  \begin{center}
    \includegraphics[width=1.0\textwidth]{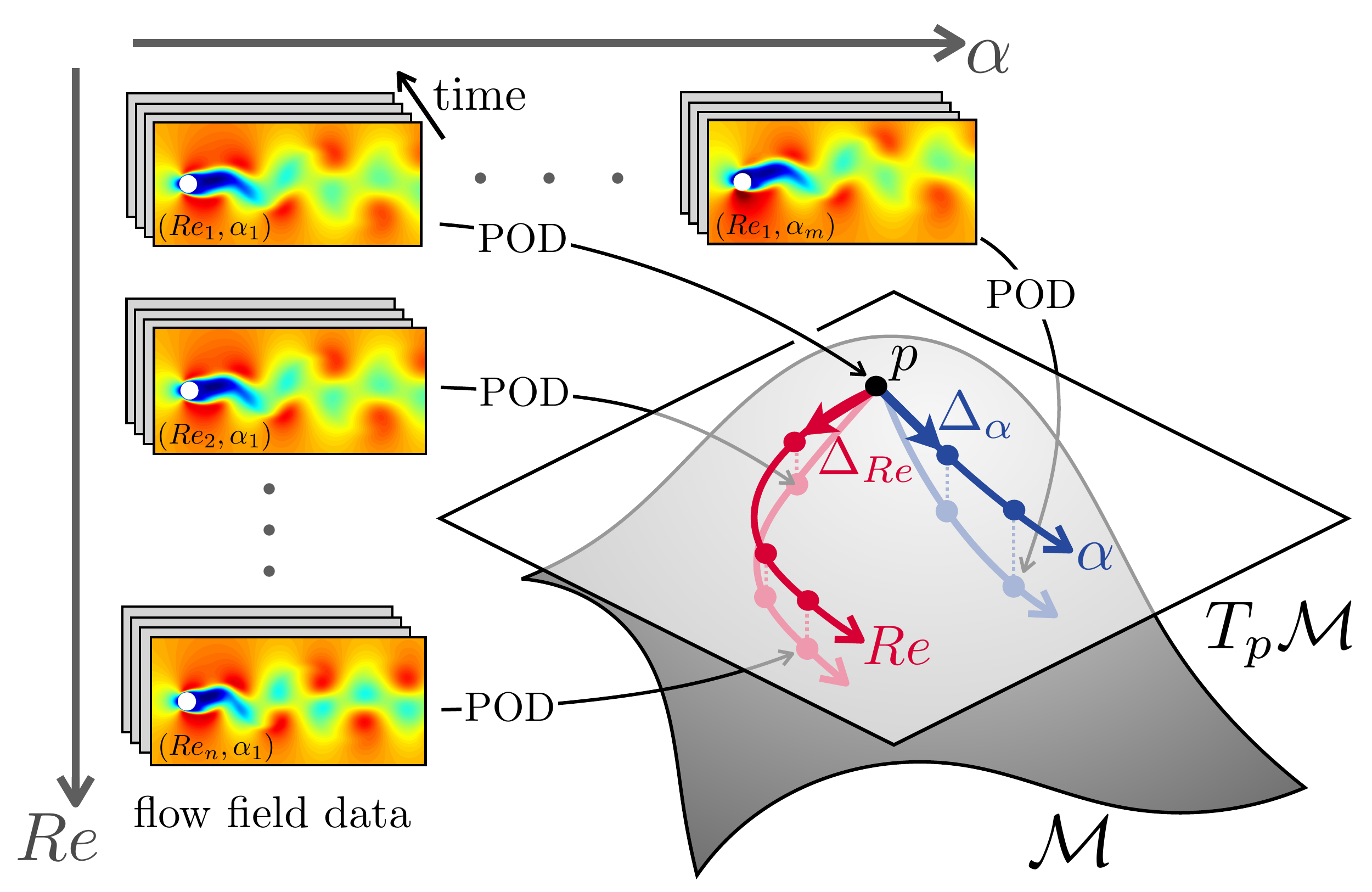}
    \caption{Schematic of the representation of sets of POD modes extracted from the flow field dataset in a wide range of flow parameters in terms of a matrix manifold $\mathcal{M}$.
      The variation of the characteristic structures of the fluid flow around a rotating cylinder with the Reynolds number $Re$ and rotation rate $\alpha$ are described as curves on $\mathcal{M}$.
      The relationship between the matrix manifold $\mathcal{M}$ and a tangent vector space at $p$, represented as $T_{p}\mathcal{M}$, is also described.
      Tangent vectors in the tangent-vector space in the Reynolds-number and rotation-rate directions are represented as $\Delta_{Re}$ and $\Delta_{\alpha}$, respectively.\label{fig:MManifold}}
  \end{center}
\end{figure}
Figure~\ref{fig:MManifold} shows a schematic representation of the sets of POD modes extracted from the flow field dataset
over a wide range of flow parameters in terms of a matrix manifold.
We consider the flow field around a rotating cylinder, where the flow parameters are the Reynolds number and rotation rate of the cylinder.
The POD modes are extracted from the time series of the flow field data obtained from an experiment or numerical simulation for each flow parameter.
A single matrix represents a set of POD modes for each flow parameter.
The set of these matrices forms a subset of a matrix manifold and is the subject of the analysis in this study.
The relationship between the two matrices can be defined by considering a matrix manifold and introducing an appropriate Riemannian metric.
This enables the discussion of the relationship between the flow fields at different parameters through the POD modes or the subspace spanned by them.
We then consider that the set of POD modes at a specific parameter (or the subspace spanned by them) varies with the Reynolds number or rotation rate.
The variation of the POD modes (or subspace) as a function of Reynolds number or rotation rate is corresponding to tracing a curve on the matrix manifold.
The main objective of this study is to investigate the parameter dependence of fluid flows
by analyzing and modeling the curve represented by a subset of sampled sets of POD modes or subspaces.

When analyzing or modeling a subset of elements in a matrix manifold (e.g., the Grassmann manifold or Stiefel manifold),
considering the tangent vector space at an element on the manifold and performing the analysis in the tangent vector space is more practical in terms of the numerical computations.
The Grassmann manifold is not a vector space because it does not admit vector addition and scalar multiplication in a natural way. 
However, the tangent vector space at each point of the Grassmann manifold has a vector-space structure.
This structure allows vector addition and scalar multiplication in the tangent vector space,
enabling the consideration of the linear interpolation of the tangent vectors.
By considering a mapping, referred to as an exponential map, each tangent vector in the tangent-vector space can be associated with an element on the matrix manifold.
Conversely, a mapping that maps an element on the matrix manifold to a vector in the tangent-vector space is referred to as a logarithmic map.
A logarithmic map is the inverse of an exponential map.

\subsection{Mode sensitivity analysis on Grassmann manifold}
First, a brief overview of the POD methods is provided with the purpose of defining some variables.
Let $\boldsymbol{q}_i=\boldsymbol{q}(t_i)\in\mathbb{R}^N$ ($i=1,2,\ldots,N_t$) be snapshot data indicating an instantaneous fluctuating flow field at a given time $t_i$,
where $N$ is the dimension of each snapshot data (the number of grid points multiplied by the number of variables) and $N_t$ is the number of snapshots.
A snapshot matrix $Q$ is defined by arranging the snapshot data as column vectors as follows:
\begin{equation}
  Q := \begin{bmatrix} \boldsymbol{q}_1 & \boldsymbol{q}_2 & \ldots & \boldsymbol{q}_{N_t} \end{bmatrix} \in\mathbb{R}^{N\times N_t}.
\end{equation}
The POD modes $\boldsymbol{\phi}_i\in\mathbb{R}^N$ are obtained by solving the following eigenvalue problem:
\begin{equation}
  QQ^T\boldsymbol{\phi}_i = \lambda_i\boldsymbol{\phi}_i,~~~(i=1,2,\ldots,r),
\end{equation}  
where the superscript $T$ denotes the transpose and $r$ is the number of POD modes to be employed.
In practice, the method of snapshots \citep{Sirovich1987} is typically employed when the size of $N$ is large.
The POD-mode matrix $U$ is defined by arranging the POD modes as follows:
\begin{equation}
  U := \begin{bmatrix} \boldsymbol{\phi}_1 & \boldsymbol{\phi}_2 & \ldots & \boldsymbol{\phi}_r \end{bmatrix} \in \mathbb{R}^{N\times r}.\label{eq:MatDef}
\end{equation}
Note that the POD modes are orthonormal under the following inner product:
\begin{equation}
  \langle \boldsymbol{\phi}_i,\boldsymbol{\phi}_j \rangle := \boldsymbol{\phi}_j^T\boldsymbol{\phi}_i = \delta_{ij}~~~(i,j=1,2,\ldots,r), 
\end{equation}
where $\delta_{ij}$ is the Kronecker delta.
This means that $U^TU=I_r$, where $I_r\in\mathbb{R}^{r\times r}$ is the identity matrix.

We then consider a subspace $\mathcal{S}$ spanned by the POD modes $\boldsymbol{\phi}_1,\boldsymbol{\phi}_2,\ldots,\boldsymbol{\phi}_r$ and represent it as $\mathcal{S}=\mathrm{span}(U)$.
As discussed in previous studies \citep{Hay2009,Sato2021}, the extracted POD modes depend on the flow parameters of the dataset;
therefore, the subspace spanned by the POD modes also depends on the flow parameters.
To discuss the dependence of the subspace on the flow parameter from the perspective of geometry, this study considers a set whose elements are subspaces, i.e., the Grassmann manifold.
The Grassmann manifold $\mathrm{Gr}(N,r)$ is defined as the set of all $r$-dimensional subspaces of $\mathbb{R}^{N}$ (see, e.g., \cite{Absil}):
\begin{equation}
  \mathrm{Gr}(N,r) := \{ \mathcal{S}\subset\mathbb{R}^{N}~|~\mathcal{S} \mathrm{~is~a~subspace}, \mathrm{dim}(\mathcal{S}) =r \}.  
\end{equation}
A subspace spanned by a set of POD modes for a given parameter is represented as a point on the Grassmann manifold.
Therefore, a point on the Grassmann manifold represents a linear subspace that can be specified using an orthogonal projector $P\in\mathbb{R}^{N\times N}$ onto $\mathcal{S}$.
The orthogonal projector onto $\mathcal{S}$ is represented as:
\begin{equation}
  P = UU^T.
\end{equation}
By using a specific orthogonal projector $P$, a point on the Grassmann manifold is uniquely specified;
however, the matrix size of the orthogonal projector is $N$-by-$N$.
Computations using these orthogonal projectors require a large memory size when the flow field data are used as snapshot data.
Therefore, an alternative approach, referred to as the orthonormal-basis perspective \citep{Edelman1998}, is considered in this study.
A subspace $\mathcal{S}$ is specified by the matrix $U$, whose columns form an orthonormal basis of $\mathcal{S}$.
This perspective requires an $N$-by-$r$ matrix for computations on the Grassmann manifold, whereas the choice of the matrix is nonunique for determining points on the Grassmann manifold.
Two different matrices, $U_1$ and $U_2$, which consist of orthonormal bases as column vectors, span the same subspace
when there exists a matrix $R\in O(r)$, where $O(r)$ is the orthogonal group in dimension $r$, such that $R$ satisfies $U_2=U_1R$.
We consider the following equivalence class:
\begin{equation}
  [U] = \{ UR~|~R\in O(r)\}.\label{eq:Equiv}
\end{equation}
Consequently, an alternative representation of the Grassmann manifold is obtained as follows:
\begin{equation}
  \mathrm{Gr}(N,r) = \{ [U]~|~U^TU=I_r \}.
\end{equation}
A point on the Grassmann manifold can be identified by specifying $[U]$ instead of the corresponding orthogonal projector.

The tangent vector space of the Grassmann manifold at point $[U]\in \mathrm{Gr}(N,r)$ is represented as
\begin{equation}
  T_{[U]}\mathrm{Gr}(N,r) = \{ \Delta\in\mathbb{R}^{N\times r}~|~U^T\Delta = 0\},
\end{equation}
where $\Delta$ denotes a tangent vector.
This study considers the following Riemannian metric:
\begin{equation}
  \langle \Delta_1,\Delta_2 \rangle_{[U]}^\mathrm{Gr} = \mathrm{trace}(\Delta_1^T\Delta_2),\label{eq:GrMetric}
\end{equation}
where $\Delta_1,\Delta_2\in T_{[U]}\mathrm{Gr}(N,r)$ (i.e., $U^T\Delta_i=0, (i=1,2)$).

The subspaces spanned by sets of the POD modes at two different parameters are represented by two different points on the Grassmann manifold.
Let us consider the shortest path between two points on the Grassmann manifold, which is referred to as a geodesic.
Let $\gamma(\xi)$ be the geodesic in the Grassmann manifold, parametrized by $\xi\in\mathbb{R}$.
There exists an open interval containing $0$ and a unique geodesic that satisfies $\gamma(0)=[U_p]\in\mathrm{Gr}(N,r)$ and $\frac{d}{d\xi}\gamma(0)=\Delta\in T_{[U_p]}\mathrm{Gr}(N,r)$ (see, e.g., \cite{Lee}).
The exponential map on the Grassmann manifold $\mathrm{Exp}_{[U_p]}^{\mathrm{Gr}}(\Delta)$ maps a tangent vector $\Delta\in T_{[U_p]}\mathrm{Gr}(N,r)$ to a point on the Grassmann manifold $\gamma(1)\in\mathrm{Gr}(N,r)$.
Hence,
\begin{equation}
  \mathrm{Exp}^{\mathrm{Gr}}_{[U_p]}: T_{[U_p]}\mathrm{Gr}(N,r) \rightarrow \mathrm{Gr}(N,r),~~~ \Delta \mapsto \gamma(1).\label{eq:GrGeodesic}
\end{equation}
The exponential map on the Grassmann manifold can be written explicitly as follows \citep{Edelman1998}:
\begin{equation}
  \mathrm{Exp}^{\mathrm{Gr}}_{[U_p]}(\Delta) = [U_pV\mathrm{cos}(\Sigma)V^T+X\mathrm{sin}(\Sigma)V^T],
\end{equation}
where
\begin{equation}
  \Delta \stackrel{\rm{SVD}}{=} X\Sigma V^T,
\end{equation}
where $\stackrel{\rm{SVD}}{=}$ indicates performing a singular value decomposition on the matrix on the left-hand side.
The tangent vector is represented by $N$-by-$r$ matrix.
The tangent {\it vector} indicates that it is an element of the tangent vector space $T_{[U]}\mathrm{Gr}(N,r)$, which is a {\it vector space}.

Let $[U_p],[U_q]\in\mathrm{Gr}(N,r)$.
We consider a neighborhood of $[U_p]$ where there is a unique tangent vector $\Delta'\in T_{[U_p]}\in\mathrm{Gr}(N,r)$ such that $\mathrm{Exp}_{[U_p]}^\mathrm{Gr}(\Delta')=[U_q]$.
The mapping that determines this tangent vector $\Delta'$ is defined in the neighborhood and is referred to as a logarithmic map:
\begin{equation}
  \mathrm{Log}_{[U_p]}^\mathrm{Gr}:\mathrm{Gr}(N,r)\rightarrow T_{[U_p]}\mathrm{Gr}(N,r),~~~[U_q] \mapsto \Delta', \label{eq:GrLog}
\end{equation}
(see \cite{Bendokat2024} for details regarding the neighborhood where the logarithmic map is defined).
The logarithmic map on the Grassmann manifold is computed as follows \citep{Amsallem2008}:
\begin{equation}
  \Delta' = X_q\mathrm{tan^{-1}}(\Sigma_q)V_q^T,
\end{equation}
where
\begin{equation}
  (I-U_pU_p^T)U_q(U_p^TU_q)^{-1}=U_q(U_p^TU_q)^{-1}-U_p \stackrel{\rm{SVD}}{=} X_q\Sigma_q V_q^T.
\end{equation}
The exponential and logarithmic maps are inverses of each other.
Therefore,
\begin{equation}
  \mathrm{Exp}_{[U_p]}^\mathrm{Gr}\circ\mathrm{Log}_{[U_p]}^{\mathrm{Gr}}([U_q]) = [U_q].\label{eq:identity}
\end{equation}
Equation~(\ref{eq:identity}) indicates that the output matrix spans the same subspace as the input matrix.
However, this does not indicate that the output matrix is identical to the input matrix \citep{Bendokat2024}.

The distance between two points on the Grassmann manifold $[U_p],~[U_q]\in\mathrm{Gr}(N,r)$ is calculated as follows \citep{Edelman1998}:
\begin{equation}
  \mathrm{dist}([U_p],[U_q]) = \sqrt{\sum_{i=1}^r\mathrm{cos}^{-1}(\sigma_i)},\label{eq:Dist}
\end{equation}
where $\sigma_i$ denotes the $i$th singular value of $U_p^TU_q$.
Note that $\mathrm{cos}^{-1}(\sigma_i)$ is the principal (or canonical) angle between the two subspaces $[U_p]$ and $[U_q]$.
For any two points on the Grassmann manifold $\mathrm{Gr}(N,r)$, the distance is bounded as follows \citep{Wong1967}:
\begin{equation}
  \mathrm{dist}([U_p],[U_q]) \le \frac{\pi\sqrt{r}}{2}.
\end{equation}
In addition to (\ref{eq:Dist}), several definitions of the distance between subspaces exist \citep{Edelman1998}.

One of the objectives of this study is to investigate the parameter dependence of subspaces
spanned by the POD modes of the flow field on the Grassmann manifold by using the definitions provided above.
Let $\xi$ be the flow parameter.
The variation in the subspace with respect to $\xi$ can be described as the motion of a point along the curve $c$ on the Grassmann manifold.
\begin{equation}
  c: \mathbb{R}\supset I_\xi \rightarrow \mathrm{Gr}(N,r),
\end{equation}
where $I_\xi$ denotes the interval in which $\xi$ is defined.
In this study, we denote the displacement along the curve $c$ due to a small change in parameter $\xi$ by $ds/d\xi$,
and define it as the {\it sensitivity of the subspace} with respect to $\xi$.
In practice, the sensitivity of the subspace is computed using two points: $[U(\xi)]$ and $[U(\xi+\Delta\xi]$:
\begin{equation}
  \frac{ds}{d\xi} \approx \frac{\mathrm{dist}\left( \left[U\left(\xi \right)  \right],\left[U\left(\xi+\Delta\xi \right) \right] \right)}{\Delta\xi}.\label{eq:SubspaceSens}
\end{equation}

\subsection{Mode sensitivity analysis on Stiefel manifold}
An $r$-dimensional linear subspace of $N$-dimensional Euclidean space is considered as a point on the Grassmann manifold $\mathrm{Gr}(N,r)$.
The parameter dependence of the subspace spanned by the POD modes is discussed based on the geometric properties of the Grassmann manifold.
On the Grassmann manifold, the focus is solely on the subspace without considering the POD modes themselves.
On the other hand, from the perspective that the contribution of each POD mode to the flow field can be quantified based on the corresponding eigenvalues,
the ordering of the modes can be considered meaningful.
However, when considering a matrix $\hat{U}=\begin{bmatrix} \boldsymbol{\phi}_r,\boldsymbol{\phi}_2,\ldots,\boldsymbol{\phi}_1 \end{bmatrix}$,
which is obtained by interchanging the first POD mode with the $r$th POD mode of matrix $U$ defined in (\ref{eq:MatDef}),
matrices $U$ and $\hat{U}$ are regarded as the same point on the Grassmann manifold.
The Stiefel manifold should be considered instead of the Grassmann manifold when focusing on the parameter dependence of the POD modes themselves.

The Stiefel manifold $\mathrm{St}(N,r)$ is a set of rectangular, column-orthonormal $N$-by-$r$ matrices \citep{Edelman1998}:
\begin{equation}
  \mathrm{St}(N,r) := \{ U\in\mathbb{R}^{N\times r}~|~U^TU=I_r \}.
\end{equation}
The Grassmann manifold is regarded as a quotient manifold of the Stiefel manifold:
two column-orthonormal matrices are equivalent if they are related by the multiplication of an orthogonal matrix $R\in O(r)$, as indicated in (\ref{eq:Equiv}).
Therefore,
\begin{equation}
  \mathrm{Gr}(N,r) = \mathrm{St}(N,r)/O(r).
\end{equation}

The tangent vector space of $\mathrm{St}(N,r)$ at $U$ is represented by
\begin{equation}
  T_U\mathrm{St}(N,r) = \{\Delta\in\mathbb{R}^{N\times r}~|~U^T\Delta = -\Delta^TU\}.
\end{equation}
That is, $U^T\Delta$ is a skew-symmetric matrix.
This study considers the following canonical metric \citep{Zimmermann2017}:
\begin{equation}
  \langle \Delta_1,\Delta_2 \rangle_{U}^\mathrm{St} = \mathrm{trace}\left( \Delta_1^T\left(I_N-\frac{1}{2}UU^T \right)\Delta_2 \right),
\end{equation}
where $\Delta_1,\Delta_2\in T_U\mathrm{St}(N,r)$ (i.e., $U^T\Delta_i=-\Delta_i^TU,(i=1,2)$).
Note that other choices for the Riemannian metrics on the Stiefel manifold exist \citep{Huper2021}.

As in the case of the Grassmann manifold, a geodesic define an exponential map on the Stiefel manifold $\mathrm{Exp}_U^\mathrm{St}(\Delta)$
that maps a tangent vector $\Delta\in T_U\mathrm{St}(N,r)$ to a point on the Stiefel manifold $\gamma(1)\in\mathrm{St}(N,r)$, where $\gamma$ is a geodesic:
\begin{equation}
  \mathrm{Exp}_U^\mathrm{St}:T_U\mathrm{St}(N,r)\rightarrow\mathrm{St}(N,r),~~~\Delta \mapsto \gamma(1).
\end{equation}
The exponential map on the Stiefel manifold can be computed using the following matrix operations:
(readers may refer to \cite{Zimmermann2017} for a more detailed discussion of exponential and logarithmic maps).
Let $U_p$ and $\Delta$ be the base point on the Stiefel manifold and tangent vector, respectively.
First, we perform QR decomposition of the following matrix:
\begin{equation}
  \left( I_N-U_pU_p^T \right)\Delta = YZ,
\end{equation}
where $Y$ and $Z$ are the orthogonal and upper triangular matrices, respectively.
We then compute the following matrix exponential:
\begin{equation}
  \begin{bmatrix}M\\N \end{bmatrix} = \mathrm{exp_m}\left( \begin{bmatrix} U_p^T\Delta & -Z^T \\ Z & 0 \end{bmatrix} \right) \begin{bmatrix} I_r \\ 0 \end{bmatrix},
\end{equation}
where $\mathrm{exp_m}(A):=\sum_{i=1}^\infty A^j/(j!)$.
Finally, by defining $U_q:=\mathrm{Exp}_{U_p}^\mathrm{St}( \Delta )\in\mathrm{St}(N,r)$, we obtain
\begin{equation}
  U_q = U_pM+YN.
\end{equation}

Let $U_p,U_q\in\mathrm{St}(N,r)$.
There is a neighborhood of $U_p$ where there is a unique vector $\Delta'\in T_{U_p}\mathrm{St}(N,r)$ such that $\mathrm{Exp}_{U_p}^\mathrm{St}(\Delta')=U_q$.
Consequently, the logarithmic map on the Stiefel manifold is defined in the neighborhood of $U_p$ as the inverse map of the exponential map:
\begin{equation}
  \mathrm{Log}_{U_p}^\mathrm{St}:\mathrm{St}(N,r)\rightarrow T_{U_p}\mathrm{St}(N,r),~~~U_q \mapsto \Delta'.
\end{equation}  
Computation of the logarithmic map on the Stiefel manifold equipped with a canonical metric requires iterative methods.
In this study, we adopt the algorithm proposed by \cite{Zimmermann2017}.

We define the {\it sensitivity of the POD modes} with respect to $\xi$ as follows:
\begin{equation}
  \begin{bmatrix} \frac{\partial \boldsymbol{\phi}_1}{\partial \xi} & \frac{\partial \boldsymbol{\phi}_2}{\partial \xi} & \ldots & \frac{\partial \boldsymbol{\phi}_r}{\partial \xi} \end{bmatrix}:=\frac{\partial U}{\partial \xi}\in T_U\mathrm{St}(N,r).\label{eq:StSens}
\end{equation}
In practice, the sensitivity of the POD modes is computed as a tangent vector on the tangent-vector space $T_U\mathrm{St}(N,r)$ using two points $U(\xi)$ and $U(\xi+\Delta\xi)$.
\begin{equation}
  \frac{\partial U}{\partial \xi} \approx \frac{1}{\Delta \xi}\mathrm{Log}_{U(\xi)}^\mathrm{St}\left(U\left( \xi+\Delta\xi \right) \right).\label{eq:PODSensitivity}
\end{equation}
Note that the matrix of the POD modes $U(\xi+\Delta\xi)$ is not approximated by $U(\xi)+(\partial U/\partial \xi)\Delta\xi$.
The matrix can be approximated using an exponential map as follows:
\begin{equation}
  U(\xi+\Delta\xi) \approx \mathrm{Exp}_{U(\xi)}^\mathrm{St}\left( \frac{\partial U}{\partial \xi}\Delta\xi \right).
\end{equation}

Equation~(\ref{eq:PODSensitivity}) represents the sensitivity of the POD modes and describes how the POD modes vary with respect to changes in parameter $\xi$.
It is important to note, however, that the {\it sensitivity of the flow field} is not described as a superposition of these POD mode sensitivities.
To gain further insight, we consider to analyze the sensitivity of the flow field with respect to parameter variations as follows.
Here, we explicitly denote the dependence of the snapshot data on the parameter $\xi$ and represent it as $Q(\xi)=\begin{bmatrix}\boldsymbol{q}_1(\xi) & \boldsymbol{q}_2(\xi) & \ldots & \boldsymbol{q}_{N_t}(\xi) \end{bmatrix}$.
Our goal is to represent $\partial{\boldsymbol{q}_i(\xi)}/\partial{\xi}~(i=1,\ldots,{N_t})$ as a superposition of several vectors,
analogous to representing the flow field as a superposition of POD modes.
First, we perform a singular value decomposition of the matrix $Q(\xi)$.
\begin{equation}
  Q(\xi) \stackrel{\rm{SVD}}{=} U(\xi)\Sigma(\xi)V^T(\xi),\label{eq:SVD}
\end{equation}
where $U(\xi)\in\mathrm{St}(N,r)$ is the POD-mode matrix, $\Sigma(\xi)\in\mathbb{R}^{r\times r}$ is a diagonal matrix with singular values on the diagonal, $V(\xi)\in\mathrm{St}(N_t,r)$ is a column-orthonormal $N_t$-by-$r$ matrix of POD expansion coefficients normalized by the corresponding singular values.
Note that $A(\xi):=\Sigma(\xi)V^T(\xi)$ is a matrix of the expansion coefficients of the POD modes.
Consider the partial derivative of (\ref{eq:SVD}) with respect to $\xi$:
\begin{equation}
  \frac{\partial Q}{\partial \xi} \approx \frac{\partial U}{\partial \xi}\Sigma V^T+U\frac{\partial \Sigma}{\partial \xi}V^T+U\Sigma\frac{\partial V^T}{\partial \xi}.\label{eq:SVDDerivative}
\end{equation}
Equation~(\ref{eq:SVDDerivative}) suggests that, in addition to the sensitivity of the POD modes, the sensitivities of matrices $\Sigma$ and $V$ are also considered for the sensitivity of the flow field.
We can discuss the contributions of the change in the POD modes, singular values, and expansion coefficients to the flow field sensitivity with respect to the flow parameter by evaluating the contribution of each term on the right-hand side of (\ref{eq:SVDDerivative}).
In this study, we assume the following relationship:
\begin{equation}
  V(\xi) \approx V^\mathrm{ref}R(\xi),~~~R(\xi)\in O(r),\label{eq:assumption}
\end{equation}
where $V^\mathrm{ref}$ is defined as the matrix $V$ when the parameter is a specified reference parameter $\xi^\mathrm{ref}$, i.e., $V^\mathrm{ref}:=V(\xi^\mathrm{ref})$.
In other words, we assume that the subspace spanned by matrix $V$ is identical, regardless of the parameters.
The validity of this assumption is discussed in Appendix \ref{Appendix1}.
Substituting (\ref{eq:assumption}) into (\ref{eq:SVDDerivative}) yields the following equation:
\begin{equation}
  \frac{\partial Q}{\partial \xi} \approx \tilde{U}A^\mathrm{ref},
\end{equation}
where
\begin{gather}
  A^\mathrm{ref}:=\Sigma^\mathrm{ref}(V^\mathrm{ref})^T,\\
  \tilde{U}:=\frac{\partial U}{\partial \xi}\Sigma R^{T}(\Sigma^\mathrm{ref})^{-1}+U\frac{\partial \Sigma}{\partial \xi}R^T(\Sigma^\mathrm{ref})^{-1}+U\Sigma\frac{\partial R^T}{\partial \xi}(\Sigma^\mathrm{ref})^{-1},\label{eq:Utilde}
\end{gather}
where $\Sigma^\mathrm{ref}:=\Sigma(\xi^\mathrm{ref})$.
We finally obtain the following representation of the sensitivity of the flow field by defining $\tilde{U}=:\begin{bmatrix} \tilde{\boldsymbol{\phi}_1} & \tilde{\boldsymbol{\phi}_2} & \ldots & \tilde{\boldsymbol{\phi}_r} \end{bmatrix}\in\mathbb{R}^{N\times r}$ and $A^\mathrm{ref}=:\begin{bmatrix}\boldsymbol{a}^\mathrm{ref}_1 & \boldsymbol{a}^\mathrm{ref}_2 & \ldots & \boldsymbol{a}^\mathrm{ref}_{N_t} \end{bmatrix} \in \mathbb{R}^{r\times N_t}$ (where $\boldsymbol{a}^\mathrm{ref}_i=\begin{bmatrix} a^\mathrm{ref}_1(t_i) & a^\mathrm{ref}_2(t_i) & \ldots & a^\mathrm{ref}_r(t_i) \end{bmatrix}^T$):
\begin{equation}
  \frac{\partial \boldsymbol{q}_i}{\partial \xi} \approx \sum_{j=1}^ra_j^\mathrm{ref}(t_i)\tilde{\boldsymbol{\phi}_j}, ~~~(i=1,\ldots,N_t).\label{eq:SensMode}
\end{equation}
Equation~(\ref{eq:SensMode}) shows that the sensitivity of the flow field represented by the superposition of $\tilde{\boldsymbol{\phi}_j}$
is similar to the representation of the flow field by the superposition of the POD modes.
Owing to the assumption of (\ref{eq:assumption}),
the spatial structure, which includes the contributions of the sensitivities of the POD modes, singular values, and expansion coefficients,
can be represented by the time-independent vector $\tilde{\boldsymbol{\phi}_j}$.
Equation~(\ref{eq:SensMode}) suggests that the sensitivity of the flow field can be represented by the vectors $\tilde{\boldsymbol{\phi}_j}$
with the same expansion coefficients used to represent the flow field by the superposition of the POD modes. 
Therefore, the spatial structure of $\tilde{\boldsymbol{\phi}_j}$ is expected to correspond to a mode that captures important features of the sensitivity of the flow field.
In this study, we refer to $\tilde{\boldsymbol{\phi}_j}$ as a {\it sensitivity mode}.
The expansion coefficients $a_j^\mathrm{ref}$ are already given;
therefore, only the calculation of the sensitivity modes are required to obtain the time evolution of the flow field sensitivity.

\subsection{Framework of parametric reduced-order modeling}\label{subsec:pROM}
A ROM based on the Galerkin projection is derived by projecting the Navier--Stokes equations onto the subspace spanned by the POD modes.
The set of POD modes is mutually orthonormal, which leads to a well-known system of ODEs for the expansion coefficients.
As the elements of the Grassmann manifold are represented as matrices whose column vectors form an orthonormal basis (orthonormal-basis perspective) in this study,
a similar ROM based on the Galerkin projection can be derived using the element of the Grassmann manifold.
This study considers a parametric-ROM framework by projecting the Navier--Stokes equations onto a locally optimal subspace according to a given flow parameter.
The locally optimal subspace is estimated using the subspace-interpolation technique on the Grassmann manifold \citep{Amsallem2008}.
The strategy considered in this study for constructing a parametric ROM consists of two steps:
(1) finding an appropriate subspace to describe the fluid flow to be reconstructed using subspace interpolation on the Grassmann manifold,
and (2) constructing a Galerkin projection-based ROM using the estimated subspace. 

Here, we estimate the subspace $\left[ U(\xi) \right]\in\mathrm{Gr}(N,r)$
using interpolation with two subspaces $\left[U(\xi_1) \right]$ and $\left[U(\xi_2) \right]$, where $\xi_1 < \xi < \xi_2$.
Interpolation is performed on the tangent space $T_{[U(\xi_1)]}\mathrm{Gr}(N,r)$.
First, we obtain the tangent vector $\Delta(\xi_2)\in T_{[U(\xi_1)]}\mathrm{Gr}(N,r)$
by mapping the subspace $\left[ U(\xi_2) \right]\in\mathrm{Gr}(N,r)$ onto the tangent space using a logarithmic map (\ref{eq:GrLog}).
Then, we consider the linear interpolation of the tangent vector:
\begin{equation}
  \Delta([U(\xi)]) = \frac{\xi-\xi_1}{\xi_2-\xi_1}\Delta([U(\xi_2)]).
\end{equation}
The tangent vector $\Delta([U(\xi_1)])$ corresponds to the origin of the tangent vector space $T_{[U(\xi_1)]}\mathrm{Gr}(N,r)$.
The interpolation methods commonly employed in vector spaces can be applied to the tangent-vector space.
For example, the interpolation of tangent vectors using the Lagrange interpolation has been reported \citep{Pawar2020}.
This study considers the linear interpolation on the tangent-vector space
because we are mainly focused on gaining insight into the relationship between the subspace distribution
from the perspective of the geometry of the Grassmann manifold and the errors in the flow field reconstruction using the developed parametric ROM,
rather than providing a sophisticated technique for tangent-vector interpolation.

The subspace for parameter $\xi$ is obtained using an exponential map:
\begin{equation}
  [U(\xi)] = \mathrm{Exp}_{[U(\xi_1)]}^\mathrm{Gr}\left( \Delta\left( \left[U\left( \xi \right) \right] \right) \right).
\end{equation}
As a result, an $N$-by-$r$ matrix $U'(\xi)=\begin{bmatrix} \boldsymbol{\phi}'_1(\xi) & \boldsymbol{\phi}'_2(\xi) & \ldots & \boldsymbol{\phi}'_r(\xi) \end{bmatrix}\in\mathrm{St}(N,r)$
where $[U'(\xi)]=[U(\xi)]$ is obtained.

The column vectors of $U'(\xi)$, which are obtained by subspace interpolation on the Grassmann manifold, do not necessarily coincide with the POD modes (i.e., in general, $U'(\xi) \ne U(\xi)$).
The POD modes for a given flow parameter are uniquely determined as the eigenvectors of the covariance matrix constructed from the snapshot data
(except for the eigenvector signs). 
By contrast, subspace interpolation on the Grassmann manifold estimates a matrix $U'(\xi)$ such that $U'(\xi)U'^T(\xi)\approx U(\xi)U^T(\xi)$,
where $U(\xi)$ denotes the POD mode matrix corresponding to a given parameter.
In this context, by considering the product of $U'(\xi)$ and the orthogonal matrix $R\in O(r)$,
it holds that $U'(\xi)R(U'(\xi)R)^T=U'(\xi)U'^T(\xi)\approx U(\xi)U^T(\xi)$.
This means that even if $U'(\xi) \ne U(\xi)$, the matrices $U'(\xi)$ and $U(\xi)$ span the same subspace.
Conversely, even when the matrix obtained via subspace interpolation on the Grassmann manifold spans the same subspace as that of the POD modes corresponding to a given parameter,
its column vectors do not necessarily coincide with the POD modes.
Nevertheless, the column vectors of the matrix $U'(\xi)$ are mutually orthonormal, and the subspace spanned by these vectors coincides with the subspace spanned by the POD modes.
Taking this into account, we refer to the column vectors of $U'(\xi)$ estimated via subspace interpolation on the Grassmann manifold as {\it pseudo-POD modes}.

We estimate the flow fields for the parameter $\xi$ using the Galerkin projection-based ROM with pseudo-POD modes.
Because the pseudo-POD modes, analogous to the POD modes, form orthonormal bases, the following ordinary differential equations can be employed as the Galerkin projection-based ROM:
\begin{equation}
  \frac{da_i(t;\xi)}{dt} = \sum_{j,k=0}^rF_{ijk}(\xi)a_j(t;\xi)a_k(t;\xi)+\sum_{j=0}^rG_{ij}(\xi)a_j(t;\xi)~~~(i=1,\ldots,r),\label{eq:ROM}
\end{equation}
where $a_i$ is the expansion coefficient of the $i$th pseudo-POD mode.
The coefficients $F_{ijk}(\xi)$ and $G_{ij}(\xi)$ can be obtained as follows:
\begin{gather}
  F_{ijk}(\xi) = -\langle\boldsymbol{\phi}'_i(\xi),(\boldsymbol{\phi}'_j(\xi)\cdot\nabla)\boldsymbol{\phi}'_k(\xi) \rangle~~~(j,k=0,1,\ldots,r),\\
  G_{ij}(\xi) = \frac{1}{Re}\langle \boldsymbol{\phi}'_i(\xi),\nabla^2\boldsymbol{\phi}'_j(\xi) \rangle~~~(j=0,1,\ldots,r).
\end{gather}
The coefficients $F_{ijk}$ and $G_{ij}$ depend on the flow parameter $\xi$
because these coefficients are obtained by projecting the Navier--Stokes equations onto the subspace that depends on the parameter.
As a result, the parametric ROM estimates appropriate coefficients $F_{ijk}$ and $G_{ij}$ according to the given subspace for the parameter of interest.
Additionally, the expansion coefficients depend on the selection of the basis for the subspace.
Therefore, the coefficients $F_{ijk}$ and $G_{ij}$ also depend on basis selection.
In particular, the values of $F_{ijk}$ and $G_{ij}$ obtained using the pseudo-POD modes do not necessarily coincide with those derived using the POD modes.
This arises from the difference in representing the snapshot data as $\boldsymbol{q}\approx \sum_{i}^ra_i\boldsymbol{\phi}_i$
or $\boldsymbol{q}\approx \sum_{i}^ra'_i\boldsymbol{\phi}'_i$, which merely reflects the difference in the coordinate system of the subspace.
Readers may refer to \citep{Noack2003,Noack2011} for a more detailed discussion on Galerkin projection-based ROM.

The vector $\boldsymbol{\phi}'(\xi)_0$ indicates the mean field; hence, $a_0=1$.
The mean field for parameter $\xi$ is estimated by linear interpolation using $\boldsymbol{\phi}'_0(\xi_1)$ and $\boldsymbol{\phi}'_0(\xi_2)$:
\begin{equation}
  \boldsymbol{\phi}'_0(\xi) = \frac{\xi-\xi_1}{\xi_2-\xi_1}\left( \boldsymbol{\phi}'_0(\xi_2)-\boldsymbol{\phi}'_0(\xi_1) \right)+\boldsymbol{\phi}'_0(\xi_1).
\end{equation}
We estimate the initial condition for (\ref{eq:ROM}) using $\boldsymbol{q}_1(\xi_1)$ and $\boldsymbol{q}_1(\xi_2)$.
The expansion coefficients are computed using $a_i(0)=\langle \boldsymbol{q}_1(\xi_j),\boldsymbol{\phi}'_i \rangle,~(j=1,2)$.
The obtained expansion coefficients are used to determine the initial condition by performing a linear interpolation with respect to the parameter $\xi$.

\section{Example of cylinder flow with varying Reynolds number and rotation rate}\label{Sec:SensAnalysis}
In this section and the subsequent one, we consider the mode sensitivity analysis and reduced-order modeling of the flow field around a rotating cylinder to demonstrate the utility of the proposed method.
The flow parameters considered in this study are the Reynolds number and rotation rate, which are determined as the boundary conditions of the Navier--Stokes equations.
The parameter dependencies of the subspace and POD modes are described on the Grassmann manifold and Stiefel manifold.

\subsection{Full-order modeling}
Two-dimensional compressible Navier--Stokes equations are considered as the governing equations for the full-order modeling of fluid flow past a cylinder.
The governing equations are solved numerically using the finite-difference method.
The sixth-order compact-difference scheme \citep{Lele1992} is used to evaluate the spatial derivatives.
An eighth-order compact filter is also used to stabilize the numerical dispersion \citep{Visbal2002}.
The lower-upper symmetric Gauss--Seidel implicit method \citep{Yoon1988} is employed for time integration.

An O-type grid is used with 331 and 301 grid points in the radial and circumferential directions, respectively.
The computational domain has a radius of $200D$, where $D$ is the cylinder diameter.
The minimum grid width in the radial direction is set to $0.01D$.
The Mach number of the inflow is set to $0.2$ to ignore compressibility.
In this study, the compressible Navier--Stokes equations are used as the governing equations of the full-order model.
However, we consider a free-stream Mach number that is sufficiently low for the compressibility effects to be negligible.
Consequently, we employ a Galerkin projection-based ROM based on the incompressible Navier--Stokes equations
by assuming that the obtained flow field data can be regarded as an incompressible flow field.

The following velocity is imposed on the surface of the cylinder:
\begin{equation}
  \boldsymbol{u}_\mathrm{w} = \boldsymbol{\Omega},
\end{equation}
where $\boldsymbol{u}_\mathrm{w}$ and $\boldsymbol{\Omega}$ are the velocity on the cylinder surface and angular velocity of the rotating cylinder, respectively.
The following rotation rate $\alpha$ is used as a parameter in this study:
\begin{equation}
  \alpha = \frac{\Omega D}{2U_\infty},
\end{equation}
where $\Omega$ and $U_\infty$ are the amplitudes of the angular and inflow velocity, respectively.
Along the far-filed boundary, free-stream conditions are imposed in this study.
\begin{figure}
  \begin{center}
    \includegraphics[width=1.0\textwidth]{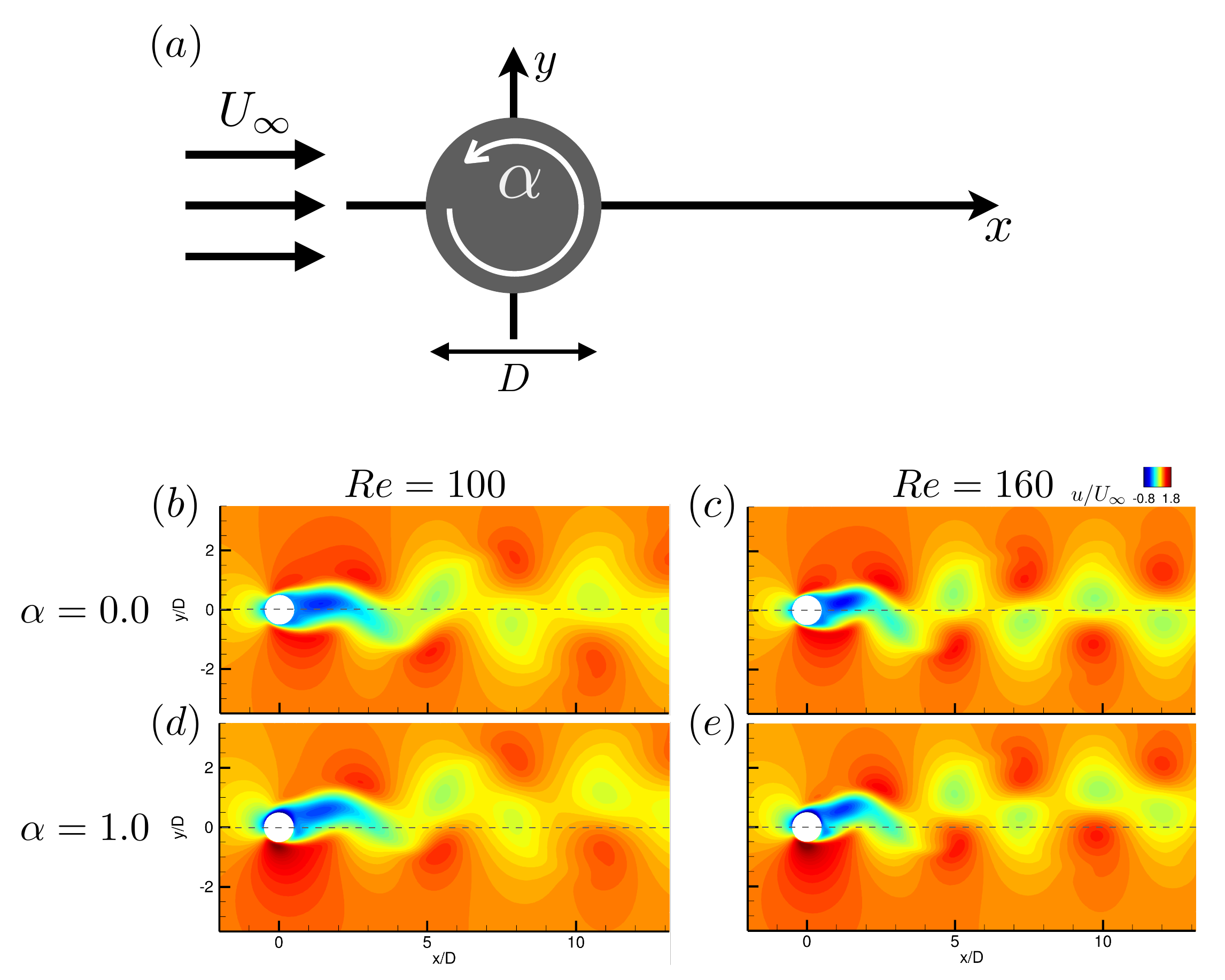}
    \caption{Schematic of the flow field to be described on matrix manifolds in this study: (a) sketch of the flow field around a rotating cylinder;
      (b,c) instantaneous spatial distributions of $x$-component velocity at $Re=100$ and $160$ without rotation; (d,e) with the rotation rate of $\alpha=1.0$.\label{fig:Schematic}}
  \end{center}
\end{figure}
A schematic of the numerical-simulation conditions is shown in Fig.~\ref{fig:Schematic}(a).

Figure~\ref{fig:Schematic}(b,c) show the instantaneous spatial distributions of the $x$-component velocity without cylinder rotation after the flow field reaches a quasi-steady state at $Re=100$ and $160$, respectively.
In both cases, a well-known K\`arm\`an vortex street can be observed.
The spacing between vortex structures in the K\`arm\`an vortex street becomes shorter as the Reynolds number increases.
The instantaneous spatial distributions of the $x$-component velocity with a rotation rate of $\alpha=1.0$ for $Re=100$ and $160$ are shown in figure~\ref{fig:Schematic}(d,e).
The K\`arm\`am vortex street is observed as in the case without rotation, and the vortex shedding is deflected upward owing to the cylinder rotation.

\begin{figure}
  \begin{center}
    \includegraphics[width=0.8\textwidth]{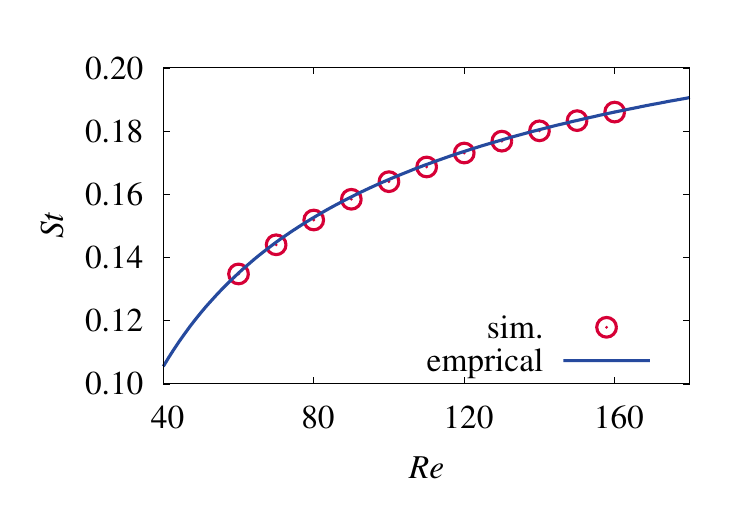}
    \caption{Comparison of the Strouhal--Reynolds number between the results obtained by the numerical simulation (circle symbol) and an empirical theory (solid line).\label{fig:StNumber}}
  \end{center}
\end{figure}
Figure~\ref{fig:StNumber} shows the Strouhal number ($St=fD/U_\infty$, where $f$ is the shedding frequency) obtained from the numerical simulation
as a function of the Reynolds number, ranging from 60 to 160.
The curve obtained from the relationship derived from the following empirical theory \citep{Williamson1998} is also shown:
\begin{equation}
  St \approx 0.2665-\frac{1.018}{\sqrt{Re}}. \label{eq:StNum}
\end{equation}
The Strouhal--Reynolds number relationship obtained by numerical simulation is in good agreement with the empirical theory.

\subsection{Sensitivity analysis on Grassmann manifold}\label{subsec:SensGr}
\begin{figure}
  \begin{center}
    \includegraphics[width=1.0\textwidth]{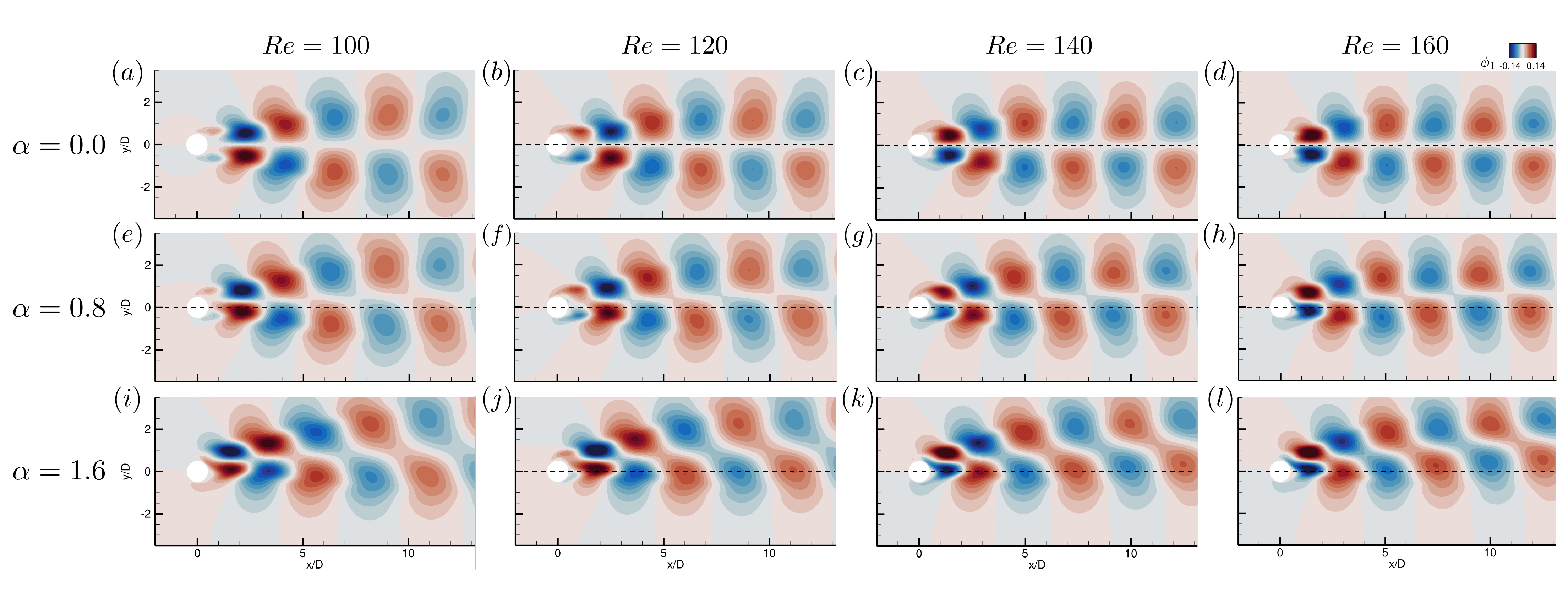}
    \caption{Spatial distributions of the first POD modes at $Re=100$, $120$, $140$, and $160$:
      (a,b,c,d) at the rotation rate of $\alpha=0.0$; (e,f,g,h) first POD modes at $\alpha=0.8$; (i,j,k,l) first POD modes at $\alpha=1.6$.\label{fig:PODModes}}
  \end{center}
\end{figure}
To extract an optimal subspace to represent the dynamics of the fluid flow in a low-dimensional space for each parameter, POD analysis was conducted.
For the POD analysis, 1000 snapshots of the spatial distributions of the $x$- and $y$-components of the velocity
obtained after the flow field reached a quasi-steady state were used.
The sampling time corresponds to ten cycles of vortex shedding, that is, the sampling-time width is $1/(100St)$.

Figure~\ref{fig:PODModes} shows the spatial distributions of the extracted first POD modes in the domain of $(Re,\alpha)\in [100,160]\times [0.0,1.6]$.
For the flow field without cylinder rotation ($\alpha=0.0$), an antisymmetric structure with respect to the $x$-axis ($y=0$) is observed downstream of the cylinder to reconstruct the K\`arm\`an vortex street, as shown in figures~\ref{fig:PODModes}(a--d).
The structure of the first POD mode varies smoothly with the Reynolds number.
At higher Reynolds numbers, the characteristic structure becomes finer.
This corresponds to the gap between the shedding vortices becoming shorter at higher Reynolds numbers.
When the cylinder is rotated at the rate of $\alpha=0.8$, a similar characteristic structure to that of non-rotating case is obtained in the first POD mode,
whereas the characteristic structure is deflected upward, as shown in figures~\ref{fig:PODModes}(e--h).
Focusing on the Reynolds-number dependence of the first POD mode at a rotation rate of $0.8$,
the characteristic structure becomes finer with an increasing Reynolds number, as in the non-rotating case.
Figures~\ref{fig:PODModes}(i--l) show the first POD modes when the rotation rate is $1.6$ for different Reynolds numbers.
The characteristic structure is further deflected upward.
Moreover, the variation in deflection angle is larger when the rotation rate is increased from $\alpha=0.8$ to $1.6$
compared to the variation from $0.0$ to $0.8$.

\begin{figure}
  \begin{center}
    \includegraphics[width=0.8\textwidth]{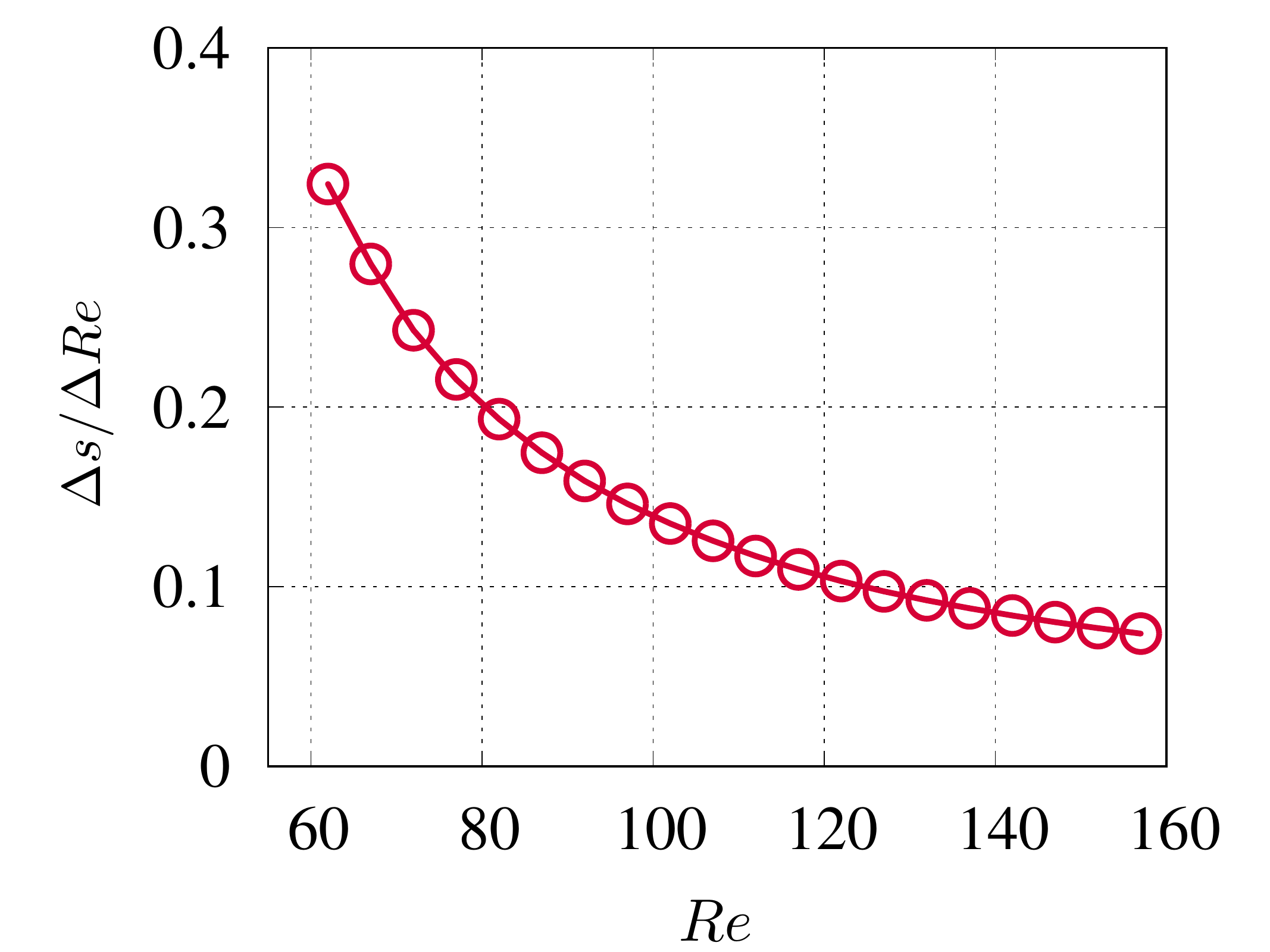}
    \caption{Sensitivity of the subspace with respect to the Reynolds-number variation as a function of the Reynolds number when the dimension of the subspace is 12.\label{fig:SubspaceSens}}
  \end{center}
\end{figure}
Figure~\ref{fig:SubspaceSens} shows the subspace sensitivity with respect to the Reynolds number obtained by (\ref{eq:SubspaceSens}) as a function of the Reynolds number
when the dimensions of the subspace are fixed at 12.
Numerical simulations using the full-order model and POD analysis are performed in the range of $Re=60$ to $160$ with $\Delta Re = 5$ to calculate the subspace sensitivities.
The sensitivity of the subspace decreases smoothly and monotonically as the Reynolds number increases.
This suggests that at lower Reynolds number, the variation of the subspace becomes larger per unit Reynolds number,
whereas the subspace variation is less significant at higher Reynolds numbers.

\begin{figure}
  \begin{center}
    \includegraphics[width=1.0\textwidth]{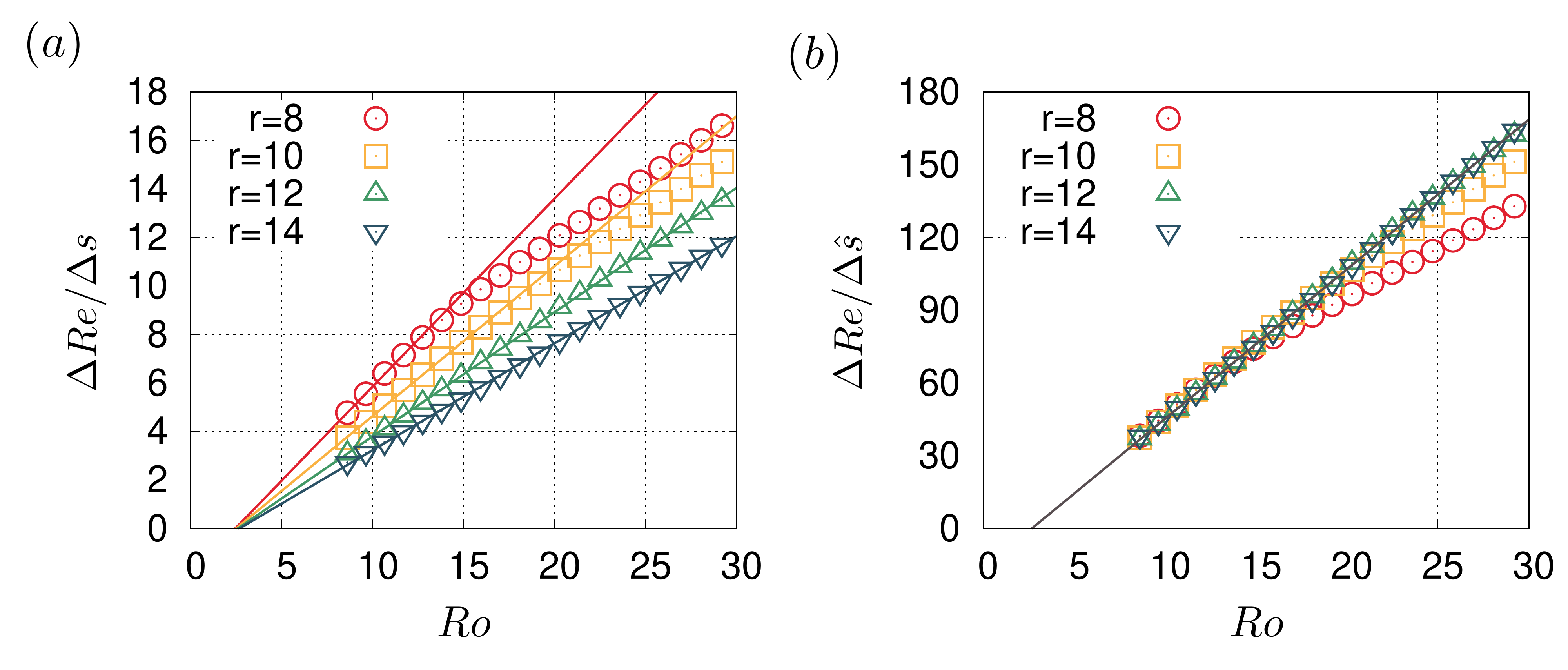}
    \caption{Relationship between the inverse of the subspace sensitivity and Roshko number $Ro(=Re\cdot St)$ for different subspace dimensions:
      (a) inverse of the subspace sensitivity calculated with (\ref{eq:SubspaceSens}); (b) inverse of the subspace sensitivity using normalized line elements by subspace dimension $\Delta\hat{s}=\Delta s/r$ instead of $\Delta s$ in (\ref{eq:SubspaceSens}). 
      Fitting curves of the form $\Delta Re/\Delta s = a\cdot Ro+b$ are also indicated.\label{fig:fig3}}
  \end{center}
\end{figure}
To discuss this in more detail, the inverse of the subspace sensitivity as a function of the Roshko number, $Ro=Re\cdot St$, for different subspace dimensions is shown in figure~\ref{fig:fig3}(a).
Fitting curves of the form $\Delta Re/\Delta s = a\cdot Ro+b$ are also plotted.
Curve fitting is performed in the ranges of $60\le Re \le 70$ and $60\le Re \le 100$ for subspace dimensions of eight and ten, respectively.
For subspace dimensions of 12 and 14, curve fitting is performed in the range of $60\le Re \le 160$.
At a low Roshko number, the inverse of the subspace sensitivity is approximately proportional to the Roshko number, regardless of the subspace dimension.
When the subspace dimension is equal to or greater than 12, the linear relationship shows good agreement with the obtained results in the range of Roshko numbers considered in this study.

Additionally, this result implies that the inverse of the subspace sensitivity becomes zero when $Ro\approx2.7$, which corresponds to $Re\approx31$ according to (\ref{eq:StNum}),
regardless of the subspace dimensions.
If we consider moving along the curve on the Grassmann manifold in the direction of the decreasing Roshko number,
figure~\ref{fig:fig3}(a) indicates that the point (subspace) corresponding to $Ro<2.7$ is inaccessible when moving along a curve from a point corresponding to $Ro>2.7$.
In other words, the subspace that characterizes the flow field for $Ro<2.7$ cannot be obtained by the continuous deformation of the subspace that characterizes the flow field considered in this study.
In view of fluid dynamics, no vortex shedding can be excited for $Re<25$, which is referred to as the diffusion dominated regime for fluid flow around a cylinder \citep{Ahlborn2002}.
\cite{Noack1994} reported that no distinct complex conjugate eigenvalue pair defining a characteristic frequency of the flow field is obtained for $Re<30$ in their stability analysis.
These results suggest that the considerable changes in the properties of the eigenvalues and eigenvectors are closely related to the bifurcation.
Our results imply that the subspace sensitivity approaches infinity when $Re\approx 31$, which corresponds to a significant change in the properties of the POD modes for the flow field around a cylinder.
This suggests the potential to detect the existence of bifurcation points by exploring the curve on the Grassmann manifold
in the direction towards which the subspace sensitivity approaches infinity.
Furthermore, the inverse of the subspace sensitivity collapses to a unified line when the line element is normalized by the subspace dimension (i.e., $\Delta\hat{s}:=\Delta s/r$), as shown in figure~\ref{fig:fig3}(b).
These results imply that the geometric features of the curve on the Grassmann manifold are closely related to the features of the fluid flow or the subspace spanned by the POD modes.
This motivates the modeling of the parameter dependence of the subspace for the dynamics of fluid flow based on the geometric features of the matrix manifolds.

\begin{figure}
  \begin{center}
    \includegraphics[width=1.0\textwidth]{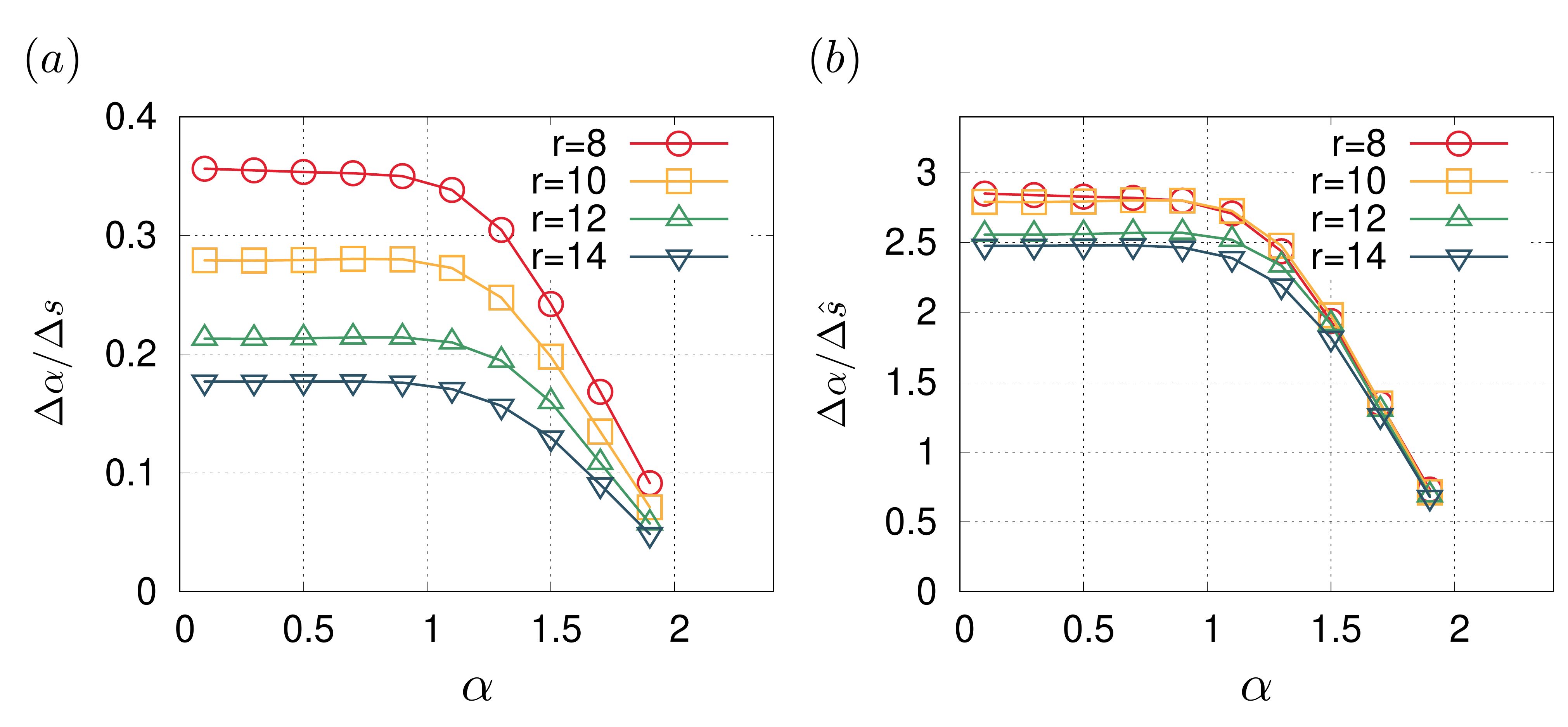}
    \caption{Relationship between the inverse of subspace sensitivity and rotation rate $\alpha$ for different subspace dimensions $r$:
      (a) inverse of the subspace sensitivity calculated with (\ref{eq:SubspaceSens}); (b) inverse of the subspace sensitivity using normalized line element. \label{fig:fig4}}
  \end{center}
\end{figure}
Figure~\ref{fig:fig4}(a) shows the inverse of the subspace sensitivity with respect to the rotation rate as a function of the rotation rate for different subspace dimensions.
The subspace sensitivity to the rotation rate is obtained using the numerical simulation of the full-order model, ranging from $\alpha=0.0$ to $2.0$ with $\Delta \alpha = 0.2$.
The subspace sensitivity is approximately constant for $\alpha<1$ regardless of the subspace dimension.
By contrast, for $\alpha>1$, the inverse of the sensitivity rapidly decreases as the rotation rate increases (hence, the subspace sensitivity increases).
This trend is consistent with the fact that the variation in the deflection angle of the first POD mode structure is larger when the rotation rate increases from
$\alpha=0.8$ to $1.6$ compared with the variation from $\alpha=0.0$ to $0.8$, as observed in figure~\ref{fig:PODModes}.

Figure~\ref{fig:fig4}(b) shows the inverse of subspace sensitivity, where the line element is normalized by the subspace dimension as a function of the rotation rate.
The inverse of the normalized subspace sensitivities for different dimensions seems to converge to a unified curve at a higher rotation rate and reaches zero
at approximately $\alpha=2.2$.
The flow past a rotating cylinder exhibits a steady state when the rotation rate increases (Hopf bifurcation)
and the critical rate is approximately $\alpha=2.0$ for $Re=100$ (e.g., \cite{Sierra2020}).
These results indicate that the inverse of the subspace sensitivity decreases to zero as the parameters approach the vicinity of the Hopf bifurcation point in both the Reynolds number and rotation-rate directions.
The points at which the inverse of the sensitivity appears to approach zero are slightly away from the Hopf-bifurcation point (on the side at which a steady solution is obtained).
This is because even in a steady flow field, the modes characterizing vortex shedding can be defined; however, these modes are stable and decay.

In addition to investigating the subspace sensitivity based on the variation in distance with the variation in the parameters,
it is useful to visualize the distribution of the subspaces for different Reynolds numbers and rotation rates to understand the parameter dependence of the subspaces.
However, visualizing the subspace distribution on the Grassmann manifold is difficult
when the dimension of the Grassmann manifold is high ($\mathrm{dim}(\mathrm{Gr}(N,r))=(N-r)r$, \cite{Absil}).
This study considers a two-dimensional visualization of the relative positions of subspaces on the Grassmann manifold for different Reynolds numbers and rotation rates.
This is achieved by using a two-dimensional polar-coordinate system defined by the norm of a tangent vector in the tangent-vector space at $(Re_0,\alpha_0)$
and by the angle, which is determined by the inner products between the two tangent vectors.
The norm of the tangent vector $\Delta\in T_{(Re_0,\alpha_0)}\mathrm{Gr}(N,r)$ is obtained from (\ref{eq:GrMetric}) as follows:
\begin{equation}
  \|\Delta \| = \sqrt{ \mathrm{trace}\left(\Delta^T\Delta \right) }.
\end{equation}
The angle between tangent vectors $\Delta_1,\Delta_2\in T_{(Re_0,\alpha_0)}\mathrm{Gr}(N,r)$ is defined as follows:
\begin{equation}
  \theta := \cos^{-1}\left(\frac{\mathrm{trace}(\Delta_1^T\Delta_2)}{\|\Delta_1\|\|\Delta_2\|}\right)~~~(0\le\theta\le \pi). \label{eq:angle}
\end{equation}
\begin{figure}
  \begin{center}
    \includegraphics[width=0.9\textwidth]{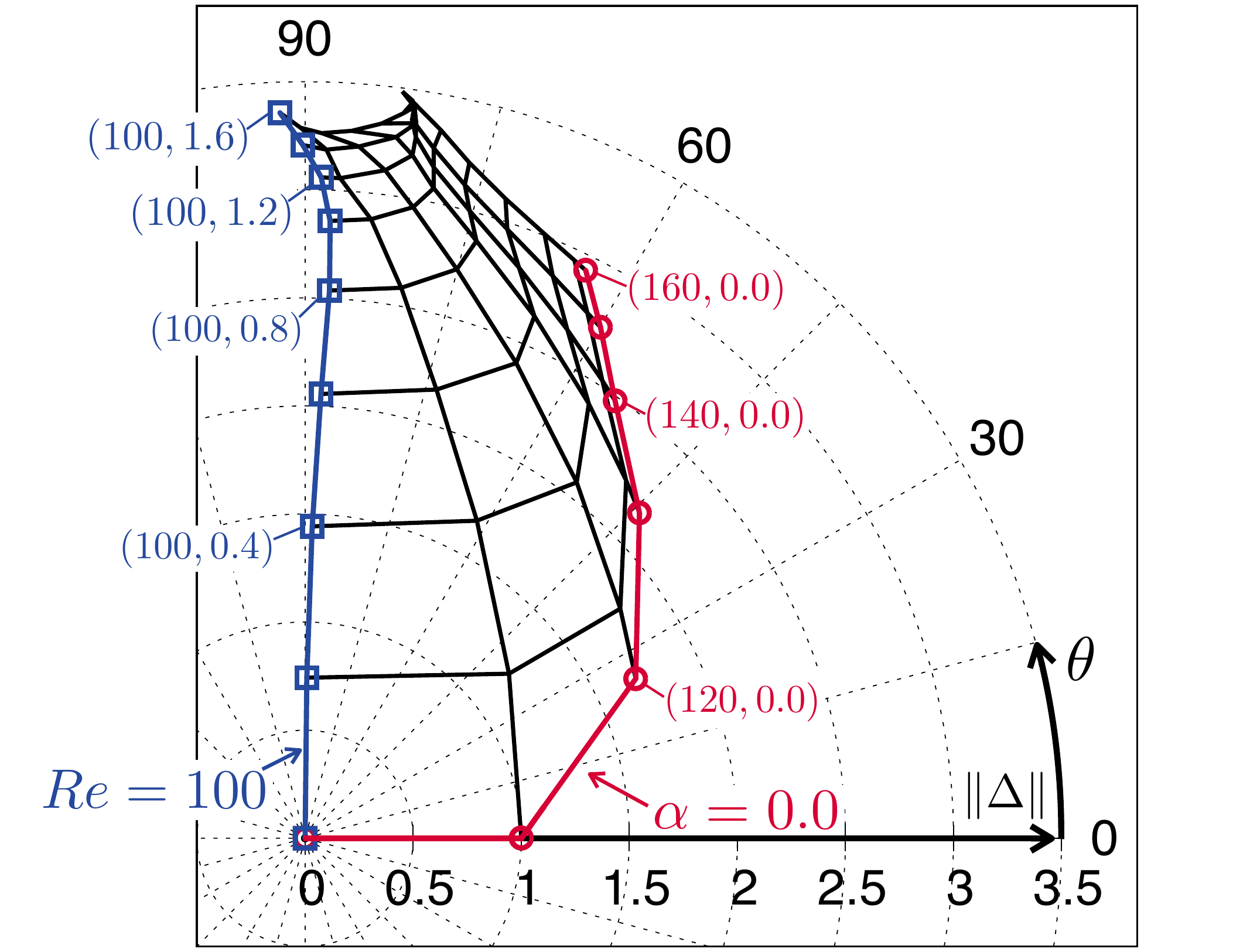}
    \caption{Subspace distribution in a domain of $(Re,\alpha)\in [100,160]\times [0.0,1.6]$ based on the norm and angle of the tangent vector in
      the tangent-vector space at $(Re,\alpha)=(100,0.0)$.
      The curves along with the Reynolds number at $\alpha=0.0$ (circle symbol) and rotation rate at $Re=100$ (square symbol) are also shown.\label{fig:fig4_2}}
  \end{center}
\end{figure}

Figure~\ref{fig:fig4_2} shows the subspace distribution in the domain $(Re,\alpha)\in [100,160]\times [0.0,1.6]$ ($\Delta Re=10,\Delta\alpha=0.2$)
based on the norm and angle of the tangent-vector space at $(Re_0,\alpha_0)=(100,0.0)$.
The norm of the tangent vector is normalized by the norm of the tangent vector for $(Re,\alpha)=(110,0.0)$.
The angle used for visualization is defined as the angle between the tangent vector for each parameter and the tangent vector for $(Re,\alpha)=(110,0.0)$.
The subspace dimension is set to 12 (hereafter, unless otherwise noted, the subspace dimension is 12).
Note that the visualization of the subspace distribution shown in figure~\ref{fig:fig4_2} depends on the tangent-vector space under consideration because the norm and angles are defined in the tangent-vector space.
The corresponding subspace is far from the tangent point when the norm is large.
The subspaces are aligned along the geodesic when the angles are constant.

First, we focus on the line describing the subspace variation in the Reynolds-number direction when $\alpha=0.0$ (circle symbol).
The angle determined by (\ref{eq:angle}) increases with the Reynolds number,
suggesting that the variation in the subspace with respect to the Reynolds number is not along a geodesic.
Instead, the subspace varies along a curved path with nonzero curvature.
In contrast, the variation in the subspace with respect to the rotation rate is relatively along with a geodesic
when the Reynolds number is fixed at 100 with a range of $0.0\le\alpha\le0.8$ (square symbol in figure~\ref{fig:fig4_2}).
The region in which the subspaces are distributed along a straight line corresponds to
the range of the rotation rate, where the subspace sensitivity in the $\alpha$-direction is constant (figure~\ref{fig:fig4}).
The angle increases with increasing rotation rate for $\alpha\ge1.0$, corresponding to the region in which the inverse of the subspace sensitivity starts to decrease.
Moreover, the angle between the tangent vectors corresponding to $(Re,\alpha)=(100,0.2)$ and $(110,0.0)$ is almost orthogonal.
This suggests that the subspace variation in the $\alpha$-direction is completely distinct from that observed in the $Re$-direction.
Consequently, visualization of the subspace distribution based on the norm and angle of the tangent vectors provides insight into the dependency of the subspace on the parameters.
This method is particularly useful for examining whether subspaces are distributed along a geodesic or a curve on the Grassmann manifold in a simple manner.

\subsection{Sensitivity analysis on Stiefel manifold}
The dependence of the subspace spanned by the POD modes on the flow parameters was previously discussed.
In this subsection, we focus on the variation in the POD modes themselves with respect to the flow parameters based on the Stiefel manifold.

\begin{figure}
  \begin{center}
    \includegraphics[width=1.0\textwidth]{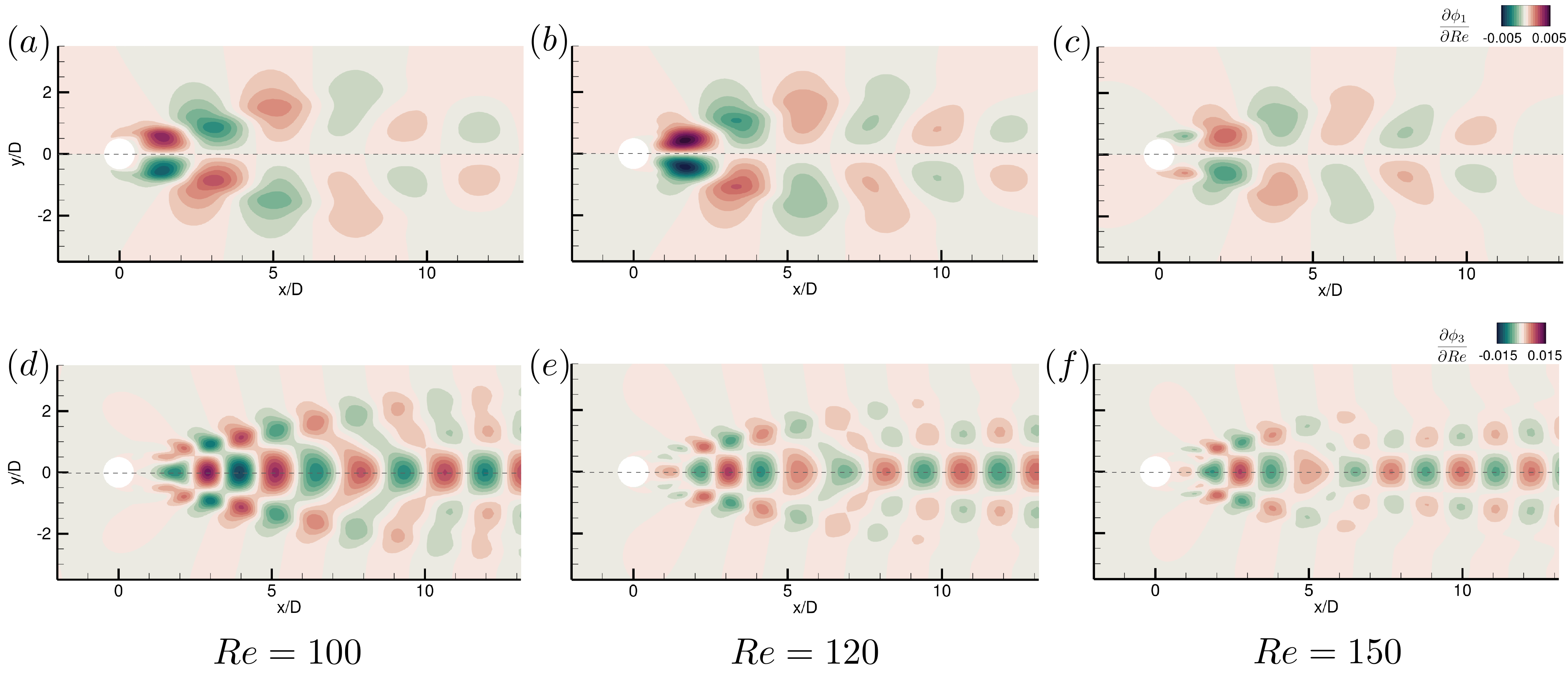}
    \caption{Spatial distributions of the sensitivity of the POD modes with respect to variation in the Reynolds number at $Re=100$, $120$, and $150$: (a,b,c) first POD modes; (d,e,f) third POD modes.\label{fig:fig5}}
  \end{center}
\end{figure}
The spatial distributions of the sensitivity of the POD modes to the Reynolds number for $\alpha=0.0$ are shown in figure~\ref{fig:fig5}.
Figure~\ref{fig:fig5}(a--c) show the sensitivity of the first POD modes, which correspond to the first column of the matrix $\partial U/\partial Re$ defined in~(\ref{eq:StSens}),
when $Re=100$, $120$, and $150$, respectively.
The sensitivities of the POD modes are evaluated based on (\ref{eq:PODSensitivity}), with $\Delta Re=10$.
The spatial distributions of the first POD-mode sensitivity exhibit an antisymmetric structure with respect to the $x$-axis, as observed in the spatial distribution of the first POD modes themselves (figure~\ref{fig:PODModes}).
These characteristic distributions of the POD-mode sensitivity represent the variation in the Reynolds number of the first POD mode,
which appears to be advected along the $x$-axis while maintaining its antisymmetric structure.
The sensitivity of the first mode for $Re=150$ indicates that the variation in the first mode is smaller than those for lower Reynolds numbers.
This corresponds to the fact that the subspace sensitivity is lower for higher Reynolds numbers, as observed in the mode sensitivity analysis of the Grassmann manifold.

The spatial distributions of the sensitivity of the third POD mode to the Reynolds number are shown in figure~\ref{fig:fig5}(d--f).
The third-mode sensitivity shows a symmetric pattern with respect to the $x$-axis, representing the advection of the third mode,
which is also characterized by a symmetric structure, downstream as the Reynolds number increases.
Similar antisymmetric and symmetric structures of the sensitivities of the first and third POD modes are observed in a previous study \citep{Hay2009}.
As observed in the first-mode sensitivity, the sensitivity of the third mode decreases slightly as the Reynolds number increases.

\begin{figure}
  \begin{center}
    \includegraphics[width=1.0\textwidth]{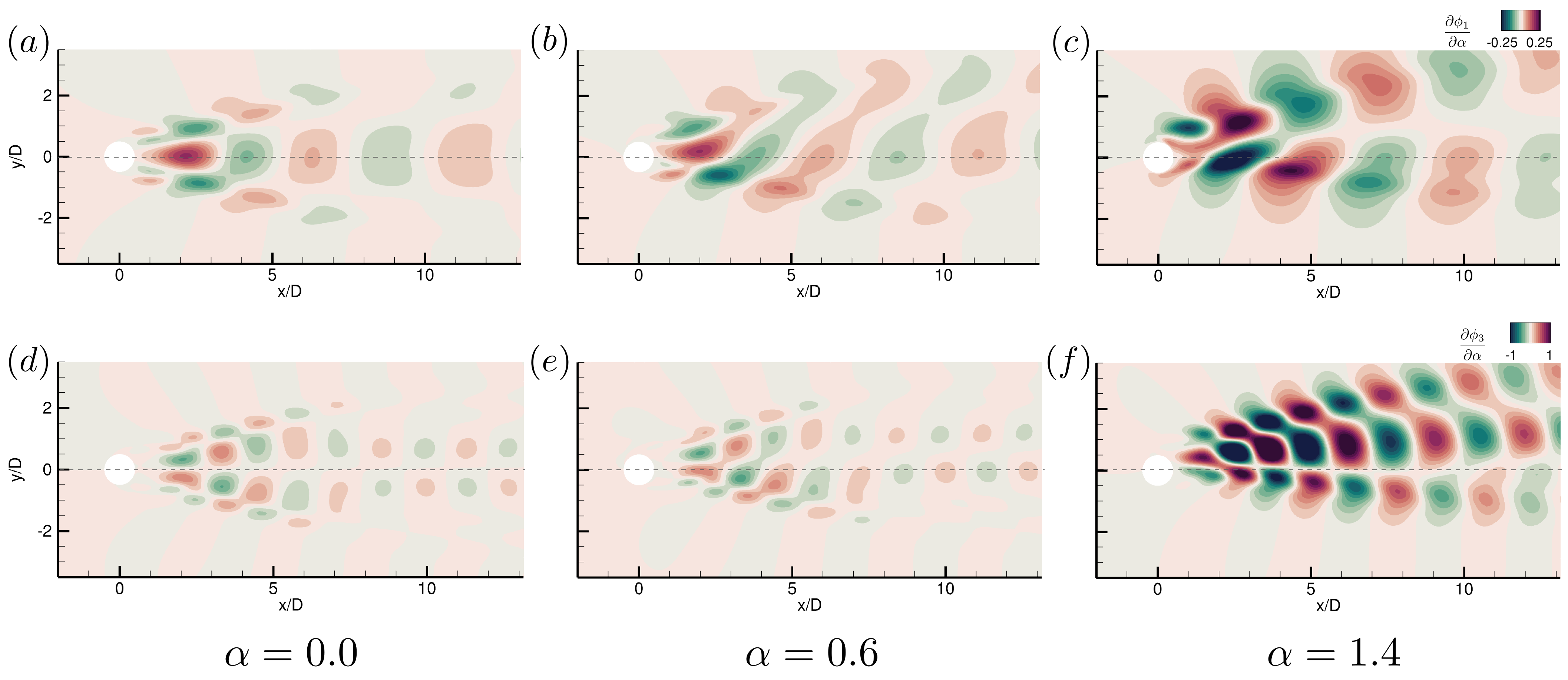}
    \caption{Spatial distributions of the sensitivity of the POD modes with respect to variation in rotation rate at $\alpha=0.0$, $0.6$, and $1.4$:
      (a,b,c) first POD modes; (d,e,f) third POD modes.\label{fig:fig6}}
  \end{center}
\end{figure}
Figure~\ref{fig:fig6} presents the spatial distributions of the sensitivities of the POD modes to the rotation rate at a fixed Reynolds number of 100.
The sensitivities of the POD modes are evaluated using $\Delta \alpha=0.2$ in (\ref{eq:PODSensitivity}).
The sensitivities of the first and third POD modes at $\alpha=0.0$ are shown in figure~\ref{fig:fig6}(a,d),
indicating approximately symmetric and antisymmetric structures with respect to the $x$-axis, respectively.
These symmetric and antisymmetric structures can be interpreted as representing the shifts along the $y$-axis of the antisymmetric and symmetric structures with respect to the $x$-axis, respectively.
The spatial distributions of the first- and third-mode sensitivities at $\alpha=0.6$ (figure~\ref{fig:fig6}b,e) show amplitudes of sensitivity similar to those at $\alpha=0.0$.
In contrast, both the first and third modes at $\alpha=1.4$ show higher sensitivities than those at $\alpha=0.0$ and $0.6$.
This indicates that the structures of these POD modes vary significantly with variations in rotation rate at $\alpha=1.4$.
These results are consistent with the finding that the subspace sensitivity to the rotation rate is approximately constant for $0.0\le\alpha\le1.0$,
while increasing for $\alpha>1.0$, as shown in figure~\ref{fig:fig4}.

\begin{figure}
  \begin{center}
    \includegraphics[width=1.0\textwidth]{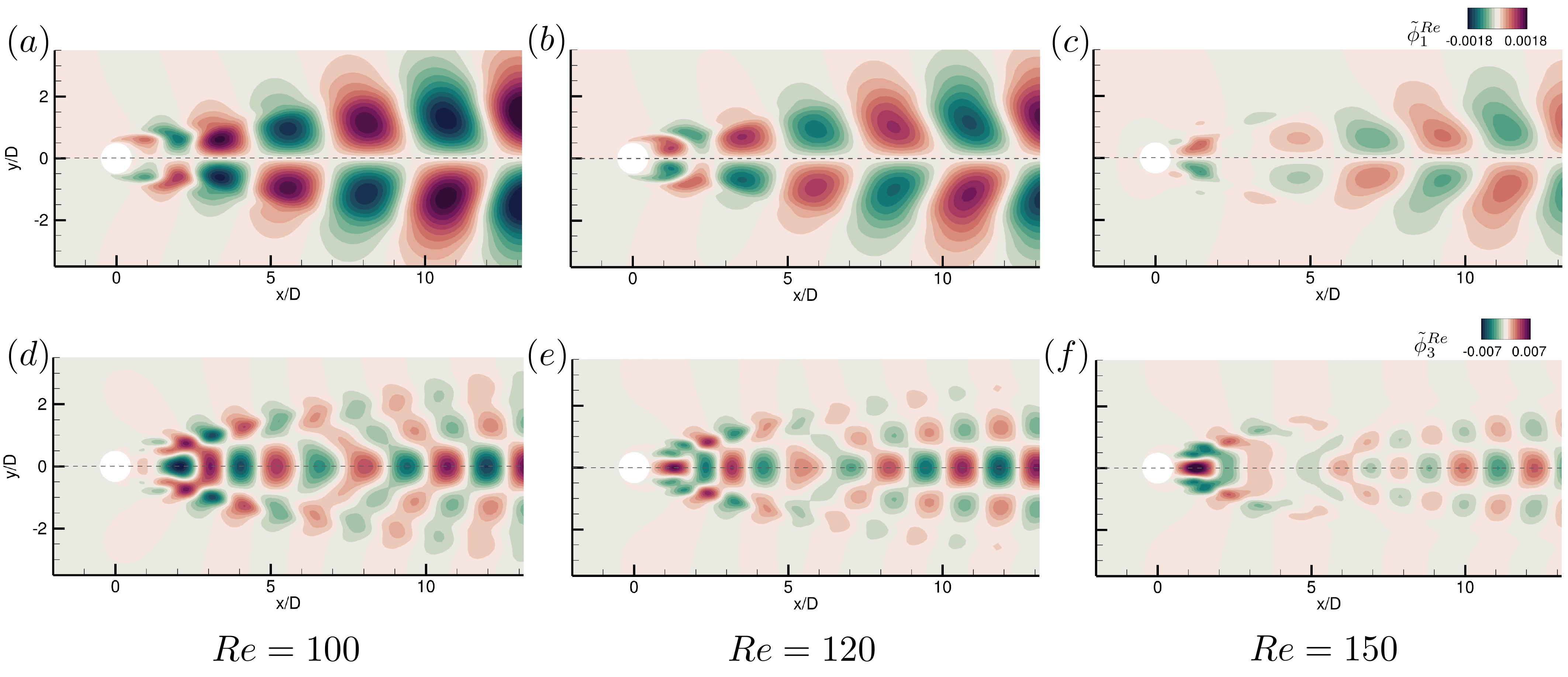}
    \caption{Spatial distributions of the sensitivity modes with respect to variation in the Reynolds number at $Re=100$, $120$, and $150$:
      (a,b,c) first sensitivity modes $\tilde{\boldsymbol{\phi}}^{Re}_1$; (d,e,f) third sensitivity modes $\tilde{\boldsymbol{\phi}}^{Re}_3$ .\label{fig:fig7}}
  \end{center}
\end{figure}
Thus far, the sensitivities of the POD modes have been analyzed.
We now focus on analyzing the flow field sensitivities using the POD-mode sensitivities defined in (\ref{eq:SensMode}).
Figure~\ref{fig:fig7}(a--c) show the spatial distribution of the first sensitivity mode with respect to the Reynolds number $\tilde{\boldsymbol{\phi}}^{Re}_1$
at $Re=100$, $120$, and $150$ when the rotation rate is fixed at $0.0$ (where superscript indicates the direction of the parameter change).
The derivative terms with respect to the Reynolds number in (\ref{eq:SVDDerivative}) are evaluated using the dataset at $Re_1=100,120,150$ and $Re_2=110,130,160$,
respectively (i.e. $\Delta Re=10$).
In this study, the reference Reynolds number is set to $Re_1$.
As observed in the first POD-mode sensitivity distributions, the spatial distributions of the first sensitivity modes exhibit an antisymmetric structure with respect to the $x$-axis
regardless of the Reynolds number.
However, the sensitivity modes exhibit different spatial patterns compared to the sensitivity of the POD modes.
This indicates that the contribution of the second and third terms in (\ref{eq:Utilde}) to the sensitivity modes is not negligible.

\begin{figure}
  \begin{center}
    \includegraphics[width=0.8\textwidth]{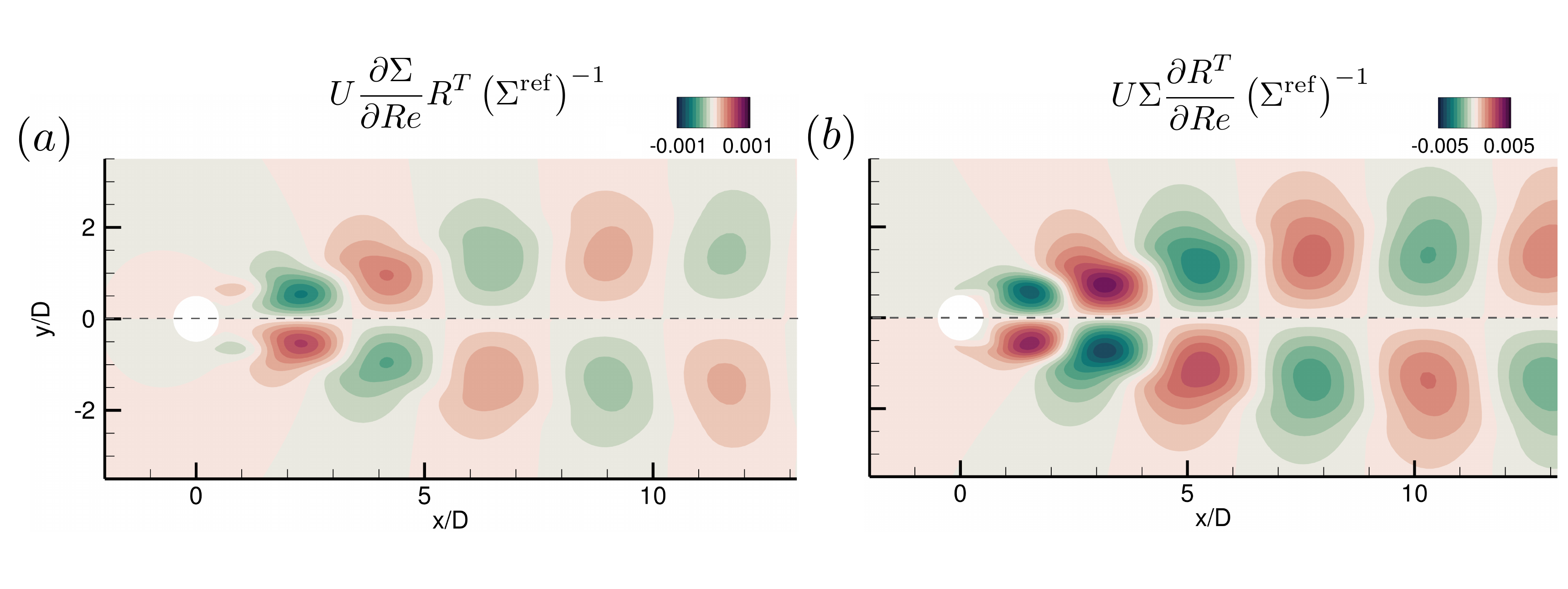}
    \caption{Spatial distributions of the contributions of the (a) second term and (b) third term in (\ref{eq:Utilde}) to the first sensitivity mode with respect to the variation in the Reynolds number at $Re=100$.\label{fig:fig7_2}}
  \end{center}
\end{figure}
The contributions of the second and third terms in (\ref{eq:Utilde}) to the first sensitivity mode with respect to the Reynolds number
at $Re=100$ are shown in figure~\ref{fig:fig7_2}(a,b), respectively.
The contribution of the first term corresponds to the sensitivity of the POD mode, which is shown in figure~\ref{fig:fig5}(a).
The second term in (\ref{eq:Utilde}), which is associated with the sensitivity of singular values to variations in the Reynolds number,
has smaller amplitudes than the first and third terms,
resulting in a slight effect on the sensitivity mode.
By contrast, the third term, which is related to the sensitivity of the semi-orthogonal matrix $V$, has an amplitude comparable to that of the first term.
This suggests that the sensitivity of the POD modes and the sensitivity of matrix $V$ are necessary to represent the sensitivity modes.
The sensitivity of $V$ is evaluated using the sensitivity of the matrix $R(Re)$, as indicated in (\ref{eq:Utilde}).
Matrix $R(Re)$ represents the phase shift of the trajectory of the expansion coefficients (normalized by the singular values) owing to the variation in the Reynolds number
(see Appendix~\ref{Appendix1} for details).
Therefore, the third term in (\ref{eq:Utilde}) can be interpreted as representing the sensitivity of the flow field caused by the phase shift of the trajectory of the expansion coefficients owing to the Reynolds-number variation.

The first sensitivity mode clearly shows that flow sensitivity decreases as the Reynolds number increases.
In particular, figure~\ref{fig:fig7}(c) indicates that at $Re=150$, the variation in the Reynolds number results in a smaller variation in the flow field compared to
the sensitivity at $Re=100$.
The third sensitivity mode also suggests that the magnitude of the flow field variations owing to the variation in the Reynolds number decreases as the Reynolds number increases,
as shown in figure~\ref{fig:fig7}(d--f).

\begin{figure}
  \begin{center}
    \includegraphics[width=1.0\textwidth]{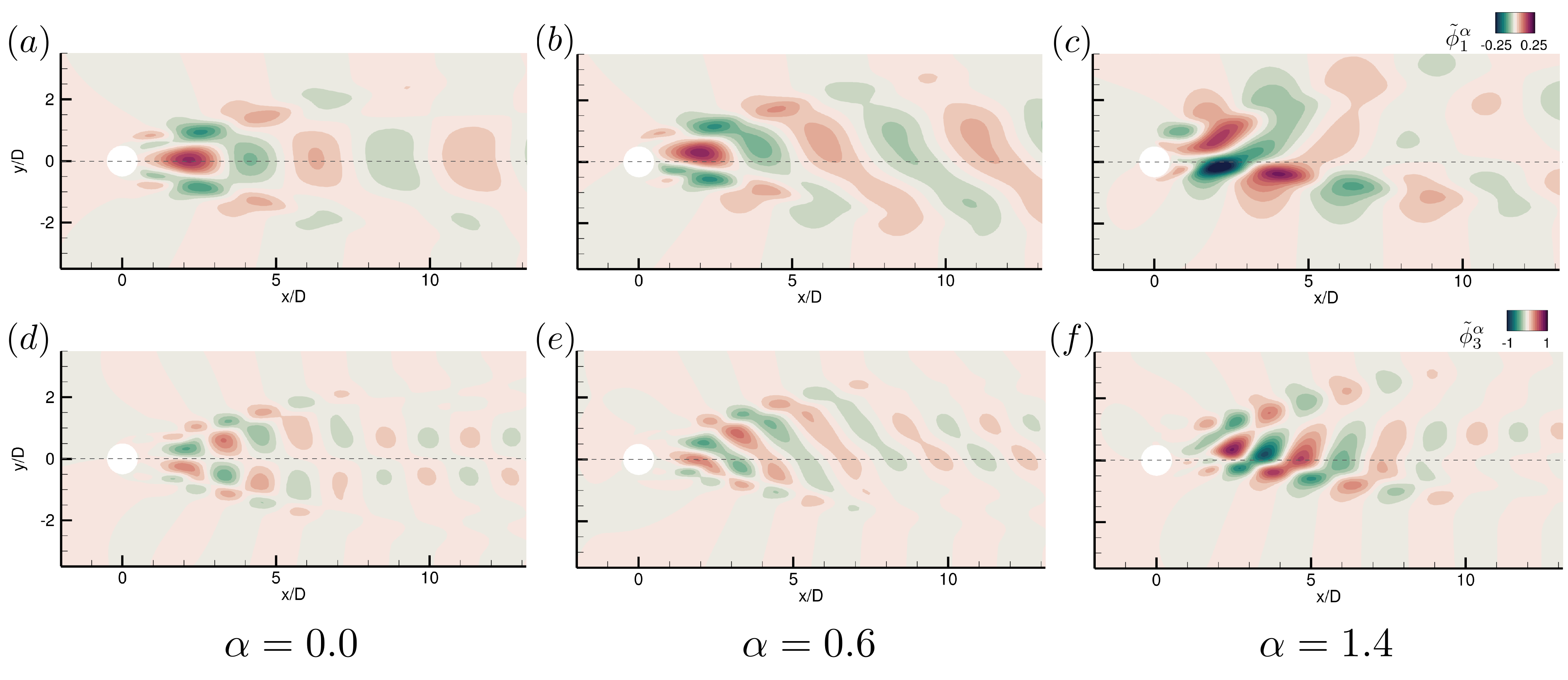}
    \caption{Spatial distributions of the sensitivity modes with respect to the variation in the rotation rate at $\alpha=0.0$, $0.6$, and $1.4$:
      (a,b,c) first sensitivity modes $\tilde{\phi}^{\alpha}_1$; (d,e,f) third sensitivity modes $\tilde{\phi}^{\alpha}_3$. \label{fig:fig8}}
  \end{center}
\end{figure}
The spatial distributions of the sensitivity modes with respect to the rotation rate at $\alpha=0.0$, $0.6$, and $1.4$ when the Reynolds number is fixed at 100 are shown in figure~\ref{fig:fig8}.
The derivative terms are evaluated using the datasets at $\alpha_1=0.0,0.6,1.4$ and $\alpha_2=0.2,0.8,1.6$ (i.e., $\Delta\alpha=0.2$).
The reference rotation rate is set to $\alpha_1$.
The first and third sensitivity modes at $\alpha=0.0$ present structures similar to the spatial distributions of the sensitivities of the first and third POD modes (figure~\ref{fig:fig8}a,d).
This indicates that the effect of the sensitivity of the POD modes is predominant compared with the second and third terms in (\ref{eq:Utilde}).
As the rotation rate increases, the structures of the first and third sensitivity modes evolve into spatial structure that differ from the spatial structures of the corresponding POD-mode sensitivities. 
In particular, the sensitivity modes at $\alpha=1.4$ suggest that the effects of the second and third terms in (\ref{eq:Utilde}) become significant, as shown in figure~\ref{fig:fig8}(c,f).

\begin{figure}
  \begin{center}
    \includegraphics[width=1.0\textwidth]{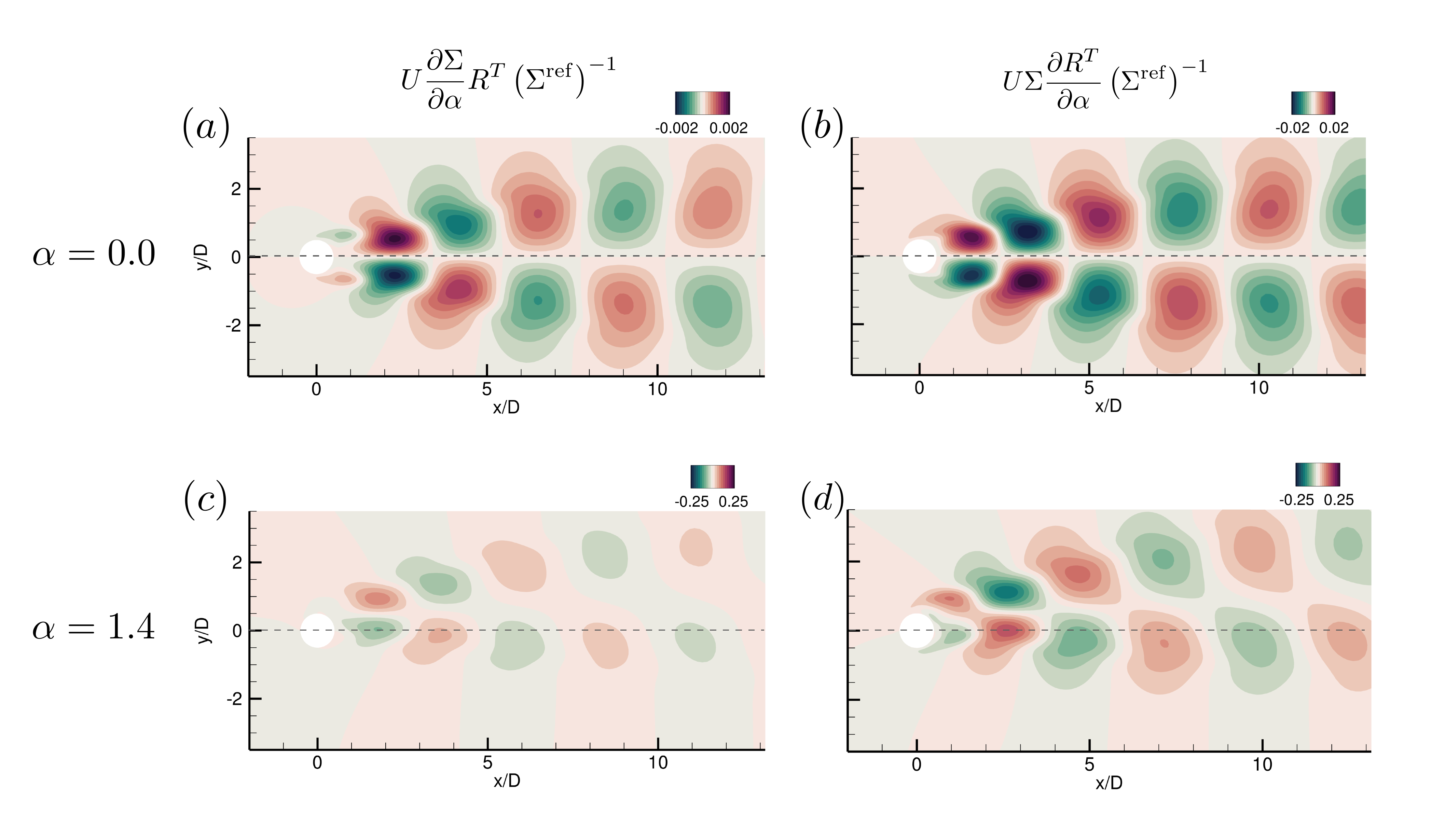}
    \caption{Spatial distributions of the contributions of (a,c) second term and (b,d) third term in (\ref{eq:Utilde}) to the first sensitivity mode with respect to variation in the rotation rate: (a,b) at $\alpha=0.0$; (c,d) at $\alpha=1.4$.\label{fig:fig8_2}}
  \end{center}
\end{figure}
Figure~\ref{fig:fig8_2}(a,b) show the contribution of the second and third terms to the first sensitivity mode with respect to the rotation rate at $\alpha=0.0$, respectively.
In contrast to the symmetric structure of the first term, an antisymmetric structure appears in both the second and third terms.
However, the amplitudes of the second and third terms are significantly smaller than those of the first term, resulting in a slight contribution to the sensitivity mode.
However, the contribution of the second and third terms are non-negligible when the rotation rate increases.
In particular, the contribution of the third term is comparable to that of the first term, as shown in figure~\ref{fig:fig8_2}.
This indicates that the sensitivity of the matrix $R(\alpha)$, which represents the phase shift of the expansion coefficient,
plays an important role in the sensitivity of the flow field, in addition to the sensitivity of the POD modes when the rotation rate is high. 

\begin{figure}
  \begin{center}
    \includegraphics[width=1.0\textwidth]{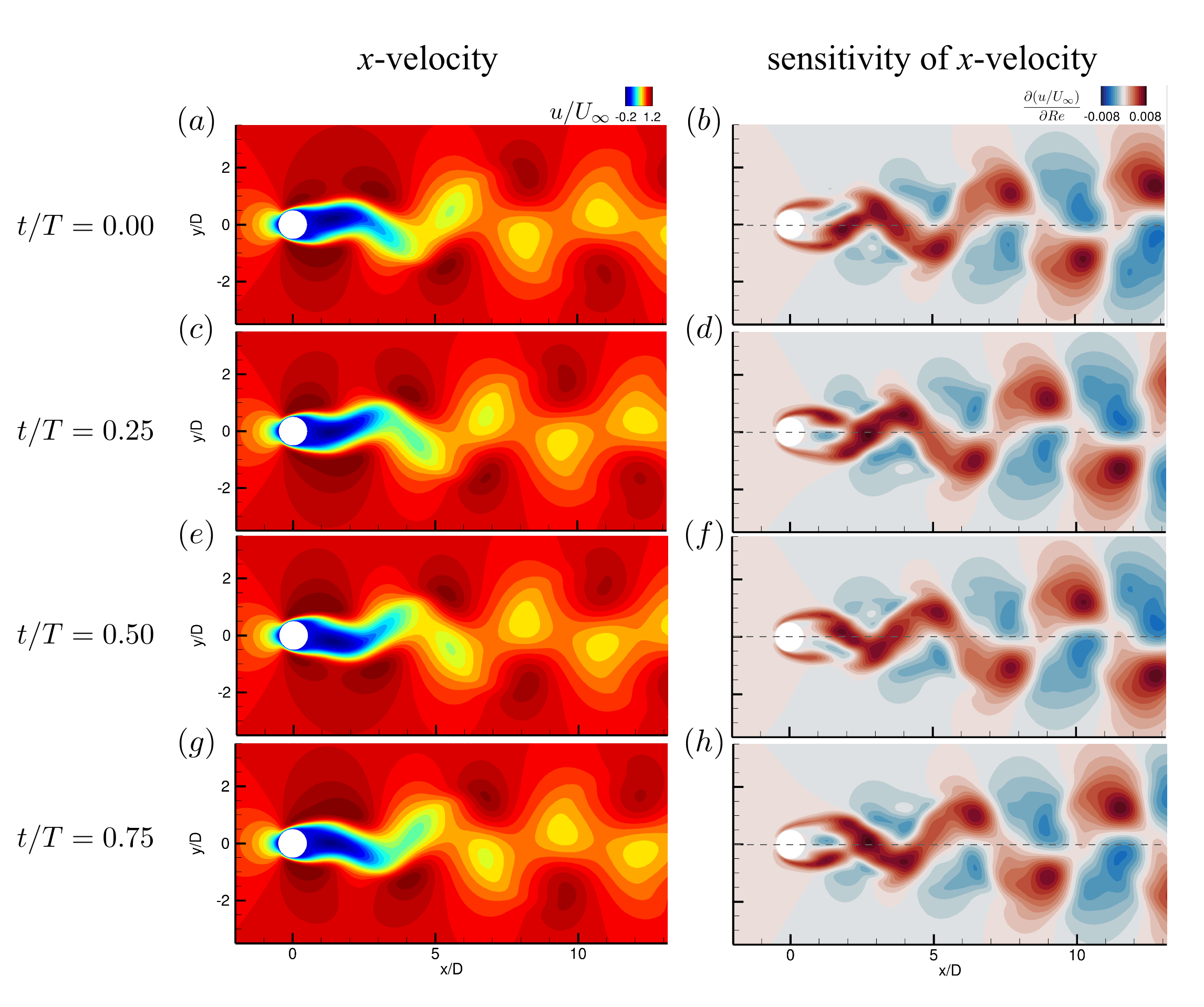}
    \caption{Spatial distributions of the flow field sensitivity with respect to variation in the Reynolds number at $(Re,\alpha)=(100,0.0)$:
      (a,c,e,g) instantaneous spatial distributions of the $x$-component velocity at $t/T=0.00$, $0.25$, $0.50$, and $0.75$;
      (b,d,f,h) distributions of the $x$-component velocity sensitivity.\label{fig:fig9}}
  \end{center}
\end{figure}
Following the above discussion of the flow field sensitivity modes,
we examine the sensitivity of the flow field by superimposing the sensitivity modes.
Figure~\ref{fig:fig9} shows the instantaneous spatial distributions of the $x$-component of the velocity and corresponding spatial distributions of its sensitivity to the Reynolds number at $(Re,\alpha)=(100,0.0)$.
Note that, in addition to (\ref{eq:SensMode}), the sensitivity distributions (figure~\ref{fig:fig9}b,d,f,h) include the sensitivity of the mean flow field.
The spatial distribution of the sensitivity at each time can be interpreted as representing how the flow field
at the corresponding time varies with small variation in the Reynolds number.
The region of high sensitivity corresponds to the areas in which the K\'arm\'an vortex street forms.
As the vortex structures advect downstream, the regions of high sensitivity advect accordingly.
Therefore, this sensitivity distribution represents the modification of the K\'arm\'an vortex street structure with the variation of the Reynolds number.
Note that the variation of the Reynolds number affects not only the flow field structures immediately behind the cylinder,
but also the structure of the flow field in the downstream region.

\begin{figure}
  \begin{center}
    \includegraphics[width=1.0\textwidth]{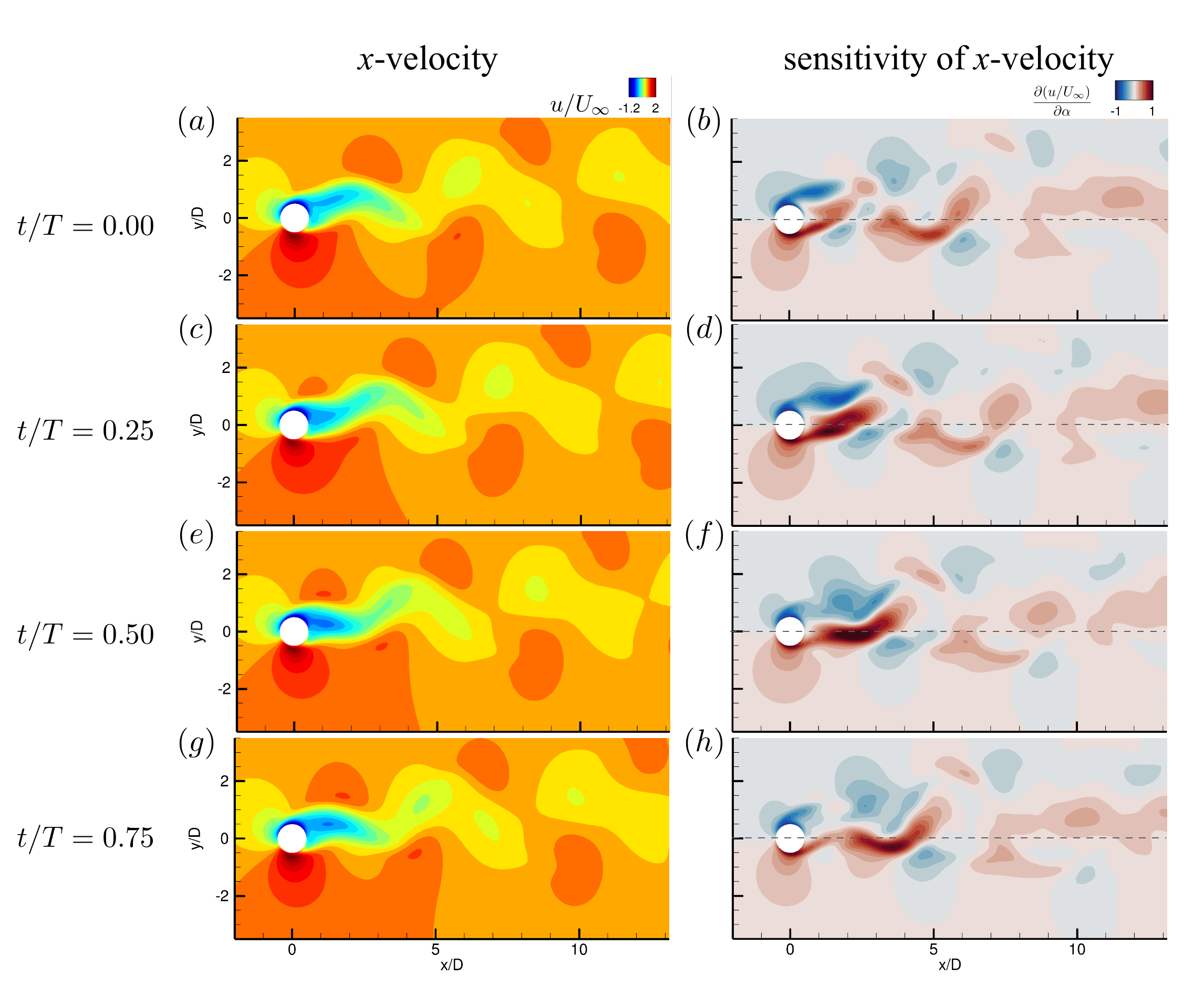}
    \caption{Spatial distributions of the flow field sensitivity with respect to variation in the rotation rate at $(Re,\alpha)=(100,1.4)$:
      (a,c,e,g) instantaneous spatial distributions of the $x$-component velocity at $t/T=0.00$, $0.25$, $0.50$, and $0.75$;
      (b,d,f,h) distributions of the $x$-component velocity sensitivity.\label{fig:fig10}}
  \end{center}
\end{figure}
Figure~\ref{fig:fig10} shows the spatial distribution of the $x$-component velocity
and corresponding spatial distribution of its sensitivity to the rotation rate at $(Re,\alpha) = (100,1.4)$.
Unlike the spatial distribution of the sensitivity to the Reynolds number,
the sensitivity distribution indicates that the variation in the rotation rate slightly affects the structures of the K\'arm\'an vortex structure.
In particular, varying the rotation rate has little effect on the flow field in the downstream region ($x/D>5$).
Instead, it indicates that the flow field has high-sensitivity regions around and immediately behind the cylinder.

As discussed above, the mode sensitivity analysis on the Stiefel manifold allows the visualization of the variation in the flow field with the flow parameters.
A distinctive feature of this method is its ability to visualize the instantaneous sensitivity of the flow field and represent the temporal evolution of the spatial distribution of the flow field sensitivity based on a superposition of sensitivity modes.
The visualization of the temporal evolution of regions, indicating high sensitivity to variations in the flow parameters, provides meaningful insights
for applications in the optimal design and active flow control of fluid machinery.

\section{Parametric reduced-order modeling}\label{Sec:pROM}
This section evaluate the performance of the parametric ROM using subspace interpolation on the Grassmann manifold.
First, we compare the reconstructed flow fields obtained by the parametric ROM, which employs POD modes estimated by direct interpolation
and the parametric ROM, which employs pseudo-POD modes obtained by the method outlined in \S{\ref{subsec:pROM}}.
We then discuss the subspace-estimation and flow-field reconstruction errors over a wide range of flow parameters when using the parametric ROM developed in this study.

\subsection{Comparison of parametric reduced-order models based on direct interpolation and subspace interpolation on Grassmann manifold}
We first discuss the performance comparison of the parametric ROMs for estimating the subspace and flow field at a given target Reynolds number $Re$ using sets of POD modes at two different Reynolds numbers $Re_1$ and $Re_2$ ($Re_1<Re<Re_2$).
In this subsection, the target Reynolds number is fixed at 90.
An intuitive direct interpolation of the POD modes to estimate the POD modes at the target Reynolds number is examined for comparison with the results obtained by subspace interpolation on the Grassmann manifold.
The direct interpolation of the POD modes is defined as the linear interpolation of the POD matrix $U(Re_1),U(Re_2)\in\mathrm{St}(N,r)$:
\begin{equation}
  U(Re) = \frac{Re-Re_1}{Re_2-Re_1}\left( U(Re_2)-U(Re_1)\right)+U(Re_1).
\end{equation}  

\begin{figure}
  \begin{center}
    \includegraphics[width=1.0\textwidth]{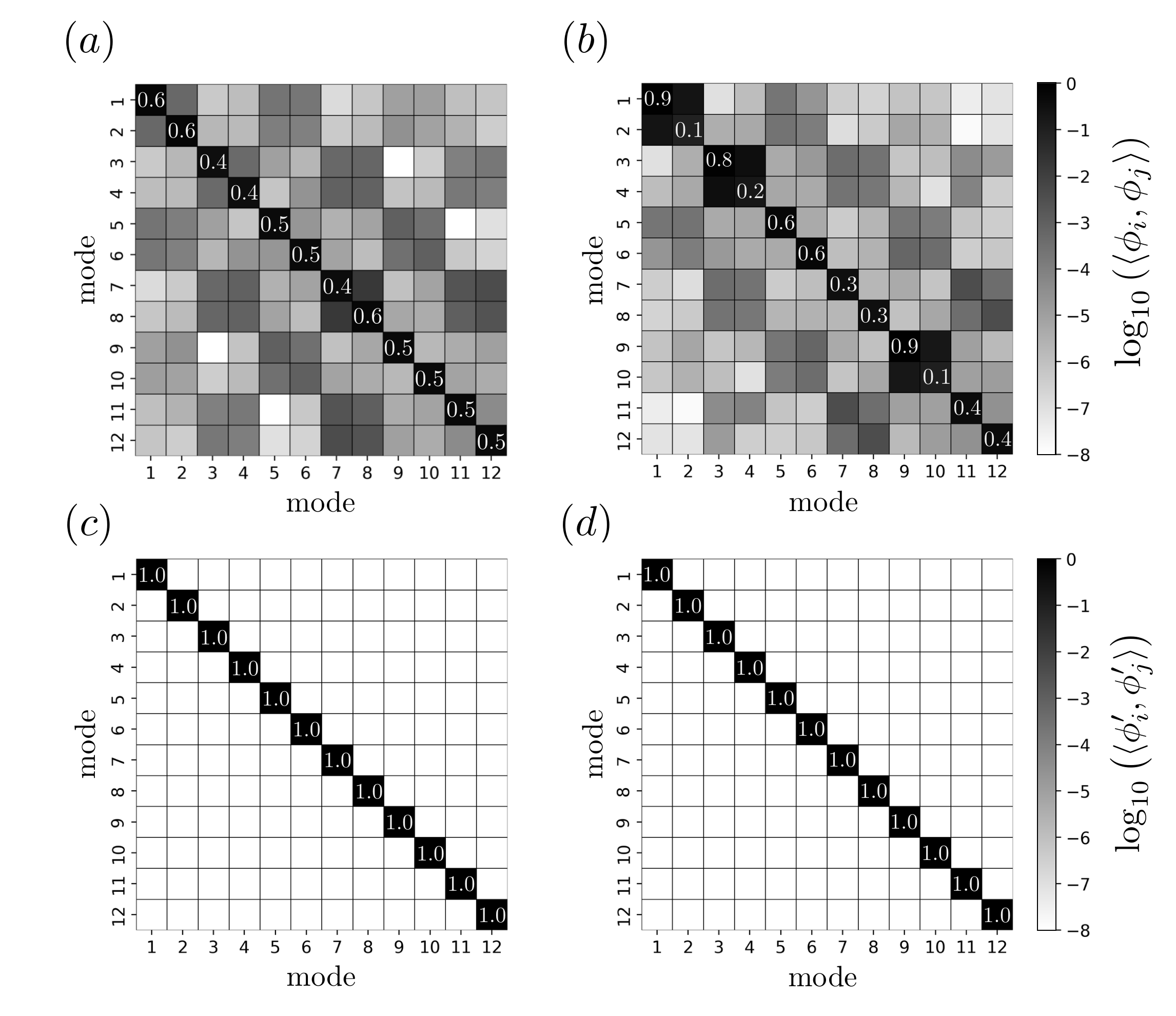}
    \caption{Inner product between the (pseudo) POD modes at $Re=90$ (a,b) estimated by direct interpolation and (c,d) subspace interpolation on the Grassmann manifold:
      (a,c) $\Delta Re=60$; (b,d) $\Delta Re=10$.
      The norms of the POD modes are represented as the diagonal components.\label{fig:fig11}}
  \end{center}
\end{figure}
Figure~\ref{fig:fig11} shows the inner product of the estimated POD modes $\langle \boldsymbol{\phi}_i,\boldsymbol{\phi}_j \rangle$.
The inner product of the POD modes obtained by the direct-interpolation method using $(Re_1,Re_2)=(60,120)$ (hereafter referred to as the $\Delta Re=60$ case) indicates that
the estimated POD modes fails to satisfy the orthonormality condition, as shown in figure~\ref{fig:fig11}(a).
The inner product of the POD modes for the case of $\Delta Re=10$ (i.e., $(Re_1,Re_2)=(85,95)$) also fails to satisfy the orthonormality condition (figure~\ref{fig:fig11}b).
This suggests that, regardless of $\Delta Re$, the set of POD modes estimated by the direct interpolation of POD modes lacks the orthonormality condition.
We can easily confirm that the directly interpolated POD set lacks orthonormality as follows:
\begin{align}
  U^T(Re)U(Re)&=\left\{ (1-c)U(Re_1)+cU(Re_2)\right\}^T\left\{ (1-c)U(Re_1)+cU(Re_2) \right\} \nonumber \\
              &=(1-2c)I_r+c(1-c)\left( U^T\left(Re_1\right)U\left(Re_2\right)+U^T\left(Re_2\right)U\left(Re_1\right) \right) \nonumber \\
              &\neq I_r,
\end{align}
where $c=(Re-Re_1)/(Re_2-Re_1)$.
This indicates that the set of POD matrices is not closed under the operation of addition,
i.e., the summation of the column-orthonormal matrices does not necessarily become a column-orthonormal matrix. 
Another clear example is the interpolation of POD matrices $U$ and $-U$.
Because $[U]=[-U]$, the result obtained by an appropriate subspace interpolation should be $[U]$.
However, direct interpolation yields a matrix in which all the POD modes are zero vectors.

Figure~\ref{fig:fig11}(c,d) show the inner product between pseudo-POD modes using subspace interpolation on the Grassmann manifold for the cases of $\Delta Re=60$ and $10$, respectively.
In contrast to the direct interpolation method, the orthonormality condition with respect to the estimated pseudo-POD modes is rigorously satisfied regardless of $\Delta Re$.
The matrix obtained by subspace interpolation on the Grassmann manifold is also an element of the Stiefel manifold,
whose elements $U\in\mathrm{St}(N,r)$ are defined as $U^TU=I_r$; that is, the column vectors are mutually orthonormal.
This property is preferable for constructing a Galerkin projection-based ROM, where the orthonormality of the (pseudo) POD modes plays an important role.

\begin{figure}
  \begin{center}
    \includegraphics[width=1.0\textwidth]{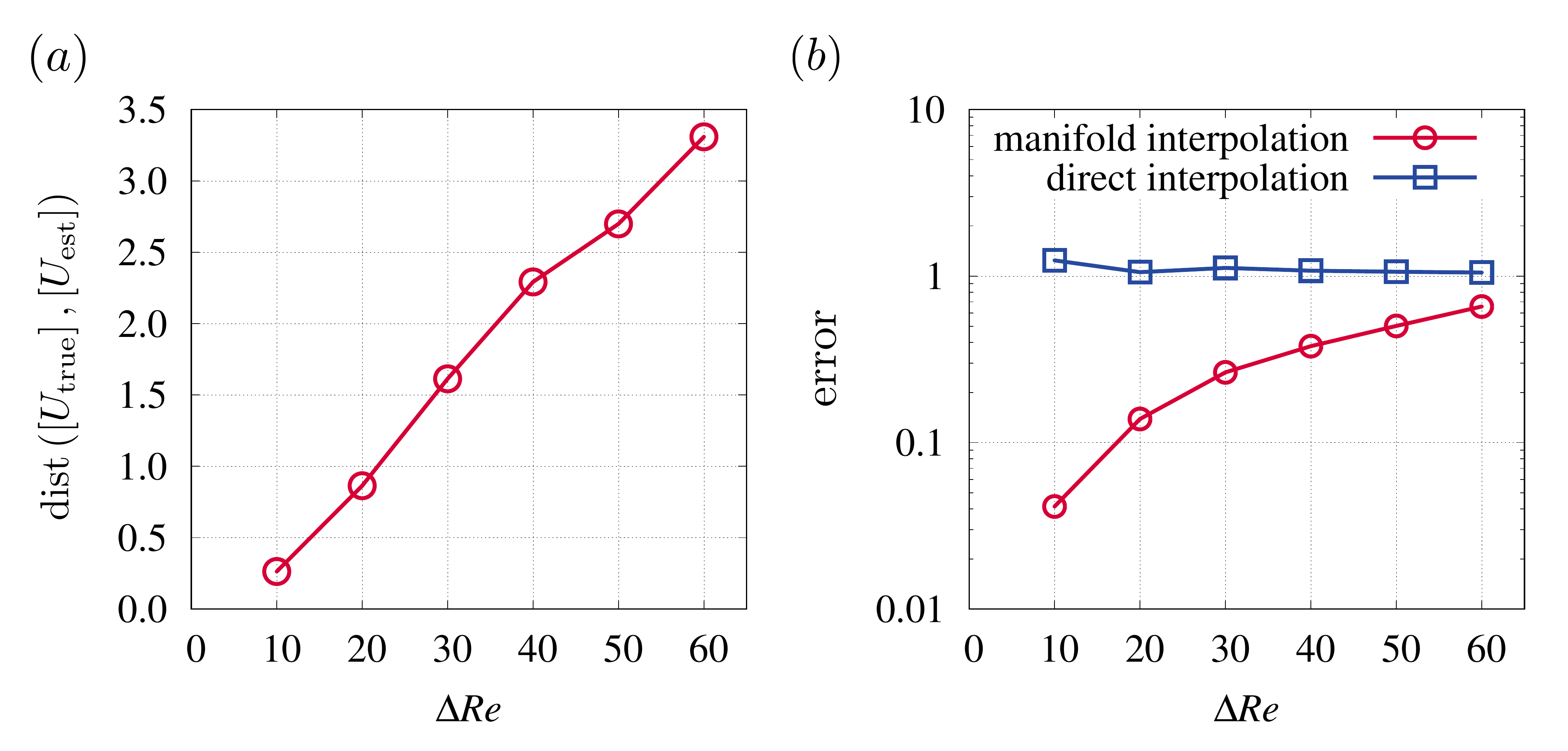}
    \caption{Error evaluation of the flow field reconstruction at $Re=90$:
      (a) subspace error estimated by the interpolation on the Grassmann manifold as a function of $\Delta Re$;
      (b) error of the flow field reconstructed by ROMs based on the subspace interpolation on the Grassmann manifold denoted as manifold interpolation (circle symbol) and direct interpolation (square symbol).\label{fig:fig13}}
  \end{center}
\end{figure}
Figure~\ref{fig:fig13}(a) shows the subspace-estimation error as a function of $\Delta Re$.
We define the subspace-estimation error as the distance between the true subspace $[U_\mathrm{true}]$, which is obtained from POD analysis using snapshot data at $Re=90$,
and estimated subspace $[U_\mathrm{est}]$.
The values of $Re_1$ and $Re_2$ are determined such that their average is $Re=90$.
The subspace-estimation error decreases approximately linearly with decreasing $\Delta Re$.
This indicates that using subspaces corresponding to Reynolds numbers in proximity to the target Reynolds number for interpolation results in convergence to the true subspace.

The errors in the flow field reconstruction obtained using the Galerkin projection-based ROM as a function of $\Delta Re$ are shown in figure~\ref{fig:fig13}(b).
These are compared with the errors obtained by the ROMs based on the POD modes estimated by the manifold and direct interpolation methods.
The error in the flow field reconstruction $\varepsilon$ is defined as follows:
\begin{equation}
  \varepsilon = \overline{\int_\mathcal{D} \frac{\| \boldsymbol{u}_\mathrm{true}(\boldsymbol{x},t)-\boldsymbol{u}_\mathrm{est}(\boldsymbol{x},t) \|_2}{\|\boldsymbol{u}_\mathrm{true}(\boldsymbol{x},t) \|_2}d\boldsymbol{x}},
\end{equation}
where $\boldsymbol{u}_\mathrm{true}$ and $\boldsymbol{u}_\mathrm{est}$ are the true and estimated velocity vectors, respectively,
and $\mathcal{D}$ denotes the simulation domain.
The bar indicates time averaging.
In this study, the ODEs for the Galerkin-projection ROM (\ref{eq:ROM}) are solved until the nondimensional time is 30.
The reconstruction error is large and almost constant regardless of $\Delta Re$ when the direct interpolation of the POD matrices is employed.
By contrast, the reconstruction error of the ROM based on the pseudo-POD modes estimated by subspace interpolation decreases with $\Delta Re$.
This indicates that reducing $\Delta Re$ used for subspace interpolation, that is, reducing the subspace estimation error, results in a reduction
in the error in the flow field reconstruction.

\begin{figure}
  \begin{center}
    \includegraphics[width=1.0\textwidth]{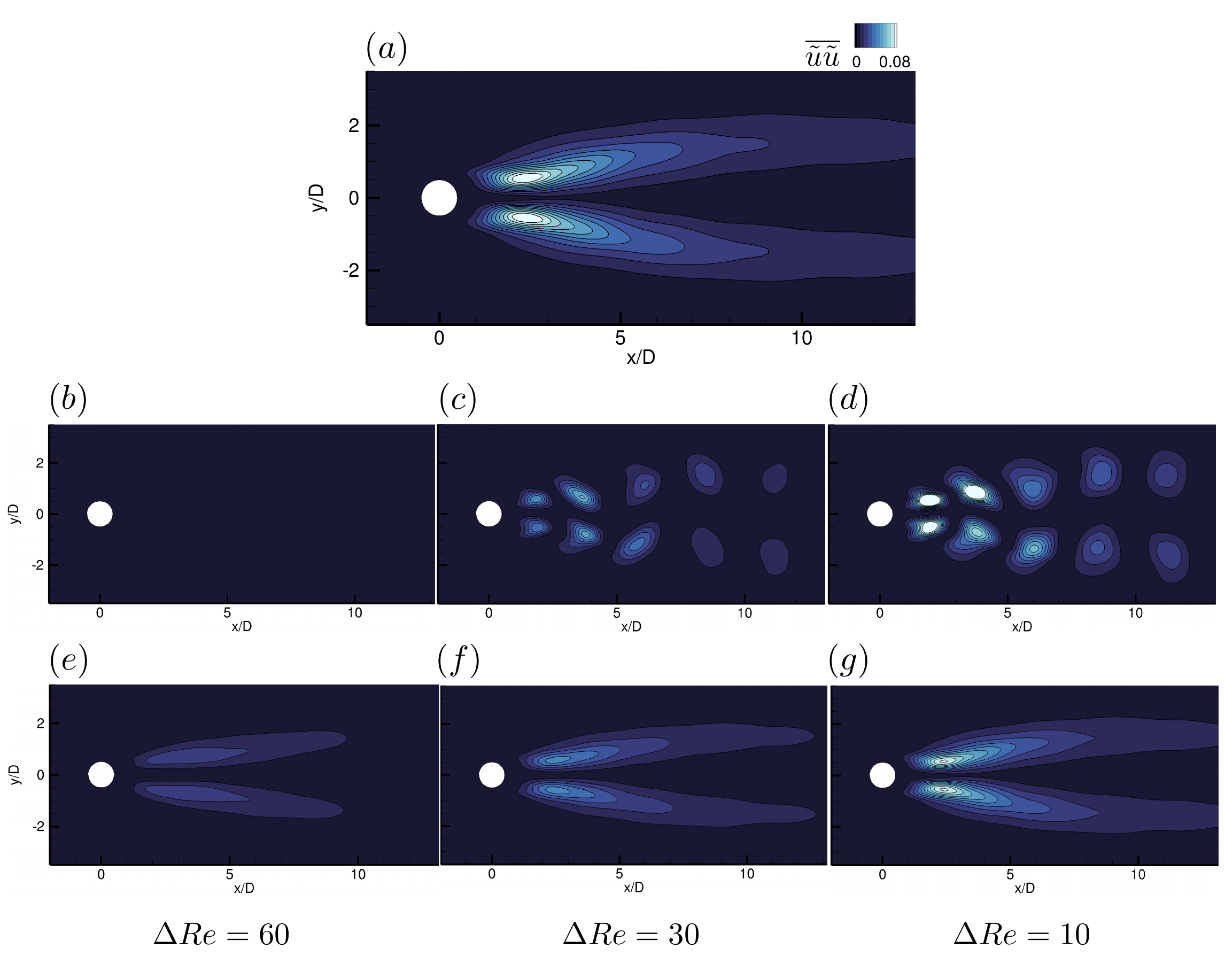}
    \caption{Comparison of the time-averaged spatial distribution of the square of the velocity fluctuation of the $x$-component:
      (a) full-order model;
      (b,c,d) ROM based on the direct interpolation of the subspace with $\Delta Re=60$, $30$, and $10$, respectively;
      (e,f,g) ROM based on the manifold interpolation.
      .\label{fig:fig14}}
  \end{center}
\end{figure}
Exploration of the spatial distribution of the flow field reconstructed by the ROM provides insights
to characterizing the property of the parametric ROM developed in this study.
Figure~\ref{fig:fig14}(a) shows the time-averaged spatial distribution of the square of the velocity fluctuation of the $x$-component obtained using the full-order model.
The spatial distributions estimated by the ROM based on direct interpolation methods for different $\Delta Re$ values are shown in figure~\ref{fig:fig14}(b--d).
These spatial distributions deviate significantly from the distribution obtained by the full-order model.
This implies that, even for a small $\Delta Re$, the reconstructed flow field is inconsistent in terms of the fluid dynamics of the flow field around a cylinder
if the subspace used in the Galerkin projection-based ROM is not appropriately estimated. 
Figure~\ref{fig:fig14}(e--g) show the spatial distributions estimated by the ROM based on subspace interpolation on the Grassmann manifold,
indicating that the spatial distribution approaches the distribution obtained by the full-order model with decreasing $\Delta Re$.
Notably, although the spatial distribution of the square of the velocity fluctuation of the $x$-component for a large $\Delta Re$ quantitatively deviates from the results of the full-order model,
the spatial structure remains qualitatively consistent.
This suggests that the subspace estimated by interpolation on the Grassmann manifold retains consistent properties in terms of fluid dynamics.

\subsection{Parametric reduced-order modeling of flow field around a rotating cylinder and error evaluations of subspace estimation and flow field reconstruction}
\begin{figure}
  \begin{center}
    \includegraphics[width=0.7\textwidth]{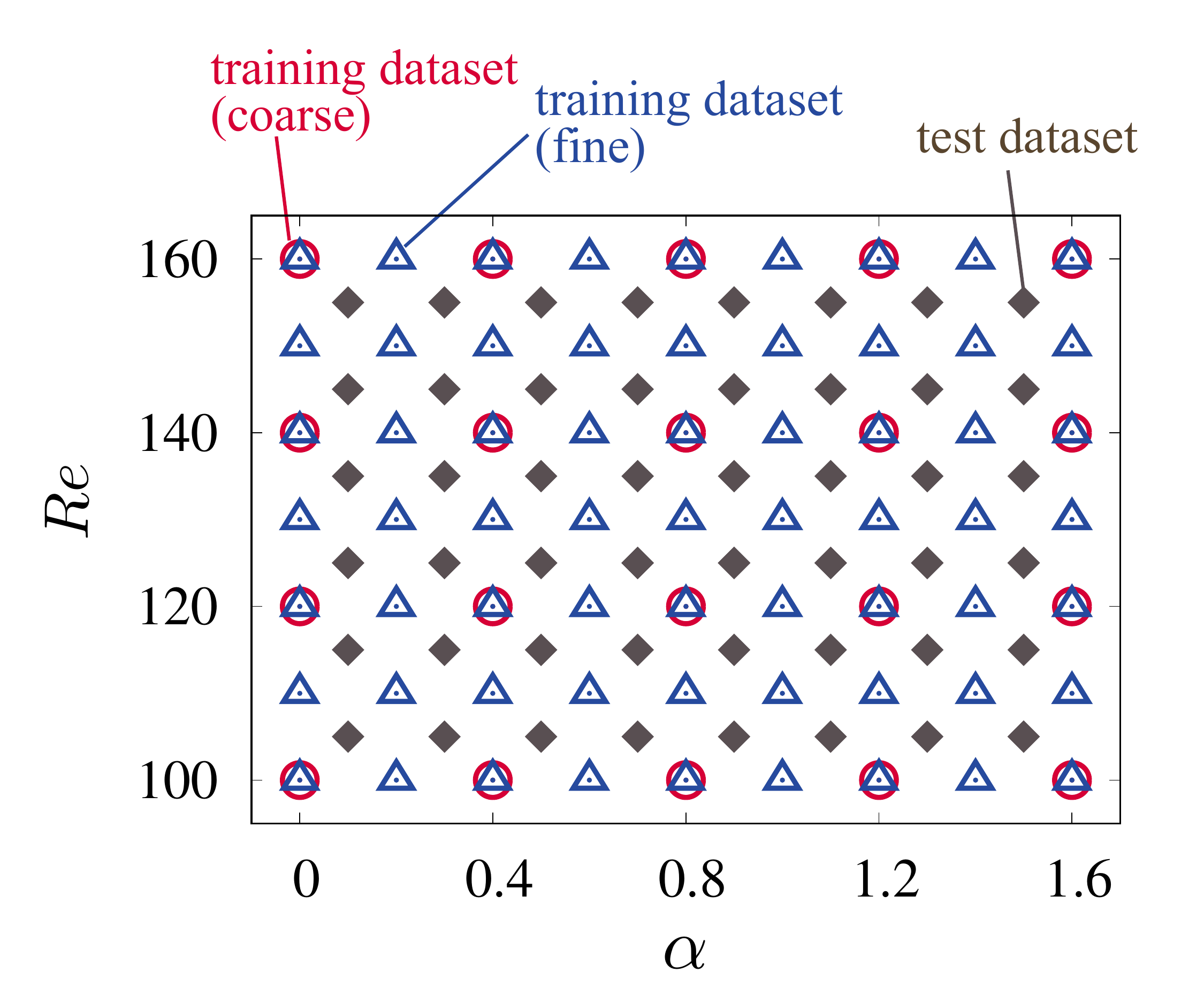}
    \caption{Sampling points $(Re,\alpha)$ used as the (circle symbol) coarse and (triangle symbol) fine training datasets, and (diamond symbol) test-data set
      for evaluation of the errors of parametric ROM in two-dimensional parameter space.\label{fig:fig16}}
  \end{center}
\end{figure}
We then evaluate the performance of the parametric ROM, which estimates the locally optimal subspace and reconstructs the flow field at a given Reynolds number and cylinder-rotation rate.
Figure~\ref{fig:fig16} shows the flow parameters (a combination of the Reynolds number $Re$ and rotation rate $\alpha$) employed to interpolate the subspaces (training dataset)
and parameters used to estimate the subspaces and reconstruct the flow field (test dataset).
Two types of training datasets, coarse- and fine-sampling data sets, are used in this study.
The subspace estimation and flow field reconstruction errors are compared between the two datasets.
For the coarse dataset, the subspaces are sampled at intervals of $\Delta Re=20$ and $\Delta \alpha =0.4$.
For the fine dataset, the subspaces are sampled at intervals of $\Delta Re=10$ and $\Delta \alpha =0.2$.
The error evaluations of the subspace estimation and flow field reconstruction in Reynolds-number direction (without cylinder rotation) are discussed in detail in Appendix \ref{Appendix2}.

The parametric ROM for two-dimensional parameter space is performed in two steps, as in the one-dimensional case:
the subspace-estimation step and Galerkin projection-based ROM step.
First, four subspaces corresponding to the conditions closest to the target parameters are selected from the training dataset.
These subspaces are mapped onto the tangent-vector space using a logarithmic map (\ref{eq:GrLog}).
The subspace for the target-flow condition is estimated using bilinear interpolation in the tangent vector space.
The initial conditions and mean field required for the Galerkin projection-based ROM are also estimated using bilinear interpolation.
In the second step, using the estimated subspace, initial condition, and mean field, the ODEs (\ref{eq:ROM})
for the Galerkin projection-based ROM are used to reconstruct the flow field.
The second step follows the same procedure as that in the case of a one-dimensional parameter space.

\begin{figure}
  \begin{center}
    \includegraphics[width=1.0\textwidth]{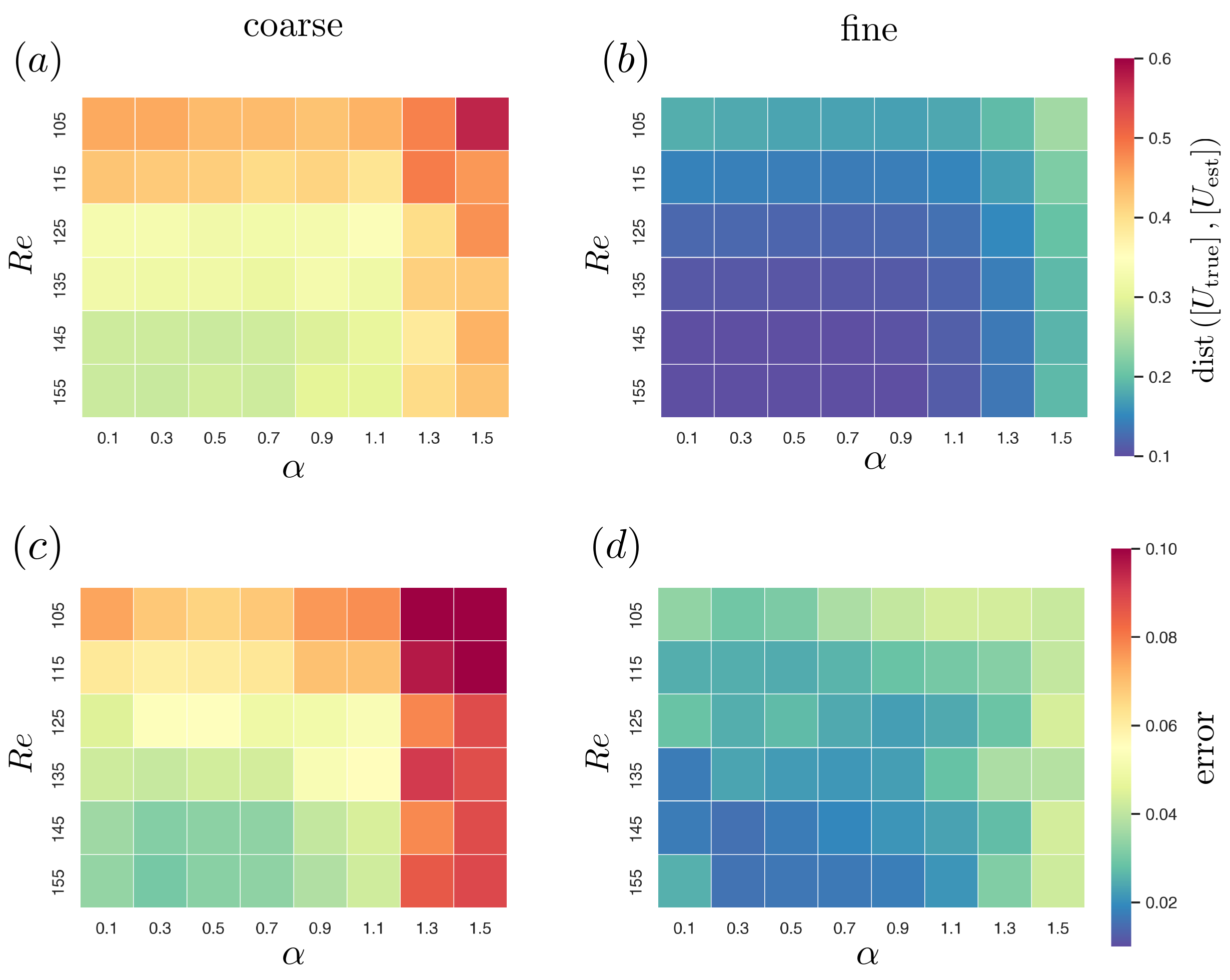}
    \caption{Error distributions of the parametric ROM in two-dimensional parameter space:
      (a,b) subspace-estimation errors when using coarse and fine training datasets, respectively;
      (c,d) flow field reconstruction errors.
      \label{fig:fig18}}
  \end{center}
\end{figure}
Figure~\ref{fig:fig18}(a,b) show the error distributions of the subspace estimation in $\alpha$-$Re$ space for the coarse and fine datasets, respectively.
The subspace-estimation error decreases as the Reynolds number increases for both the coarse and fine datasets.
This is related to the dependence of the subspace sensitivity on the Reynolds number, as discussed in \S\ref{subsec:SensGr}.
As shown in figure~\ref{fig:SubspaceSens}, the subspaces are not uniformly distributed on the Grassmann manifold when sampled at a constant interval of $\Delta Re$.
The interval between two different subspaces on the Grassmann manifold $\Delta s$ increases at lower Reynolds numbers.
Therefore, the subspace-estimation error increases as the Reynolds number decreases when the subspaces are sampled at a constant $\Delta Re$.
In addition, the error increases as the rotation rate increases, particularly in the region where $\alpha>1.1$,
which corresponds to the subspace sensitivity increasing with respect to the rotation rate for $\alpha>1.1$ (figure~\ref{fig:fig4}).
Furthermore, the estimation error decreases when the subspace is estimated using the fine dataset compared to the coarse dataset.
This indicates that reducing the sampling interval of the subspaces used for interpolation also leads to a decrease in the estimation error.

The error distributions of the flow field reconstruction by the parametric ROM using the coarse and fine datasets are shown in figure~\ref{fig:fig18}(c,d), respectively.
The flow field reconstruction error follows a trend similar to that of the subspace-estimation error:
the reconstruction error decreases as the Reynolds number increases and the rotation rate decreases.
In addition, the use of a finer dataset reduces the reconstruction error
because the use of a finer dataset reduces the subspace-estimation error.
This result suggests that, when constructing a parametric ROM, the subspace-estimation error plays a dominant role in determining the flow field reconstruction error.
This highlights that minimizing the subspace-estimation error is important for reducing the error in the flow field estimated by the parametric ROM.
A parametric ROM that reproduces flow fields with a small error over a wide range of parameters can be efficiently constructed
by sampling the subspaces finely in regions with high subspace sensitivity and coarsely in regions with low sensitivity,
because the subspace-estimation error is related to the subspace sensitivity.

\section{Conclusions}\label{Sec:Conclusions}
In this study, the sensitivity analysis of POD modes and the subspace spanned by them with respect to changes in flow parameters were analyzed
from the perspective of matrix manifolds.
This approach provides insight into how the locally optimal structures and fluid flow evolve with parameter variations,
leading to the construction of a reliable parametric ROM.
The sets of POD modes and subspaces spanned by them over a wide range of flow parameters
were represented as subsets on the Stiefel manifold and Grassmann manifold, respectively.
The relationship between the POD modes or subspaces at different parameters were analyzed using the geometric features of the curves or curved surfaces on the matrix manifolds.
The set of POD modes and subspaces were characterized by defining the Riemannian metric and distance on the Stiefel manifold and Grassmann manifold, respectively.
The results obtained from the geometric analysis on the matrix manifolds in this study provided the following insights.

First, the tangent vector along the curve on the Grassmann manifold can be interpreted as the sensitivity of the subspace.
This is closely related to the variation in the dynamics of the fluid flow due to changes in the flow parameters.
The sensitivity of the subspace, which is spanned by the POD modes for the flow field around a cylinder, increased with decreasing Reynolds number.
Our results indicate that the inverse of the subspace sensitivity increased linearly with the Roshko number, especially for higher subspace dimensions.
This relationship collapsed into a unified line when the line element of the curve on the Grassmann manifold was normalized by the subspace dimension.
In addition, the results obtained in this study indicate that the Reynolds number, at which the inverse of the subspace sensitivity becomes zero, was in good agreement with
the lower bound of Reynolds number, where the characteristic frequency of the K\'arm\'an vortex street exists (see figure~\ref{fig:fig3}).
The inverse of the subspace sensitivity with respect to the rotation rate of the cylinder decreases as the rotation rate increases
and is approaching zero as the rotation rate approaches the vicinity of the Hopf bifurcation point (see figure~\ref{fig:fig4}). 
These results imply that the flow parameter at which the subspace sensitivity approached infinity corresponded to the parameter at which the properties of the POD modes and fluid flow change significantly.
Consequently, the geometric features of the curve on the Grassmann manifold provided insights into the parameter dependence of the fluid-flow dynamics.
Additionally, the distribution of the subspaces as a function of the Reynolds number and rotation rate was visualized using the norm of the tangent vector and 
angle between tangent vectors for two different parameters.
The distribution obtained in the tangent vector space whose base point corresponded to $(Re,\alpha)=(100,0.0)$ indicated that
 the variation in the subspace to the Reynolds number is not along a geodesic.
Instead, the subspace varies along a curved path with nonzero curvature.
In contrast, the subspaces were almost aligned with a geodesic with a rotation rate in the range of $0.0\le\alpha\le0.8$.

Second, the sensitivity of the POD modes was represented as a tangent vector on the Stiefel manifold.
This enabled the analysis of the flow field sensitivity by superposing the sensitivity modes, which were defined using the tangent vector and sensitivity of the matrices
associated with the expansion coefficients.
The sensitivity of the POD modes with respect to the Reynolds number showed the displacement of the POD modes in the $x$-direction (main streamwise direction).
The sensitivity with respect to the rotation rate represented a distribution indicating the displacement of the POD modes in the $y$-direction.
The sensitivity mode of the flow field with respect to the Reynolds number exhibited a distribution similar to that of the POD mode sensitivity.
We found that not only the POD mode sensitivity, but also the expansion coefficient sensitivity played an important role in representing the flow field sensitivity.
In particular, the phase shift in the expansion coefficients due to the change in the Reynolds number had a crucial contribution to the flow field sensitivity.
Regarding the sensitivity modes with respect to the rotation rate, the sensitivity of the POD modes was predominant when the rotation rate was low,
indicating that the influence of sensitivity related to the expansion coefficient was negligible.
In contrast, when the rotation rate was high, the phase shift of the expansion coefficients affected the sensitivity of the flow field.
We also visualized the spatial distribution of the flow field sensitivity by superposing the obtained sensitivity modes.
The spatial distribution of the flow field sensitivity with respect to the Reynolds number showed how the structure of the K\'arm\'an vortex street evolved
as the Reynolds number varied, that is, the high-sensitivity region appeared along with the structure of the K\'arm\'an vortex street.
In contrast, the spatial distribution of the flow field sensitivity with the rotation rate indicates that the downstream region exhibited low sensitivity.
The flow field around and immediately behind the cylinder showed high sensitivity to changes in the rotation rate.
The spatial distribution of the flow field sensitivity with respect to the parameter changes enabled the investigations of
the variation in the flow field when deviating slightly from the design point of fluid machinery
and the effect of parameter uncertainties on the uncertainties in the flow field at a low computational cost.

Third, the subspace estimation error, which was closely related to subspace sensitivity, influences dominantly the reconstruction error of the parametric ROM based on the subspace interpolation on the Grassmann manifold.
This suggested that clarification of the parameter dependence of subspace sensitivity leads to the realization of a parametric ROM
with a small reconstruction error using a limited number of subspace samples.
The POD modes estimated using the direct interpolation method lacked orthonormality,
whereas the pseudo-POD modes estimated by subspace interpolation on the Grassmann manifold rigorously preserved orthonormality.
The parametric ROM based on direct interpolation resulted in large errors in flow field reconstruction,
even when POD modes with Reynolds numbers close to the target were used. 
In contrast, the parametric ROM using pseudo-POD modes resulted in
a decreased reconstruction error as the Reynolds-number interval for sampling the subspace decreased.
Moreover, the parametric ROM using direct interpolation yielded a flow field inconsistent with that of the full-order model regardless of $\Delta Re$. However,
the ROM based on pseudo-POD modes via subspace interpolation provided the flow field,
which showed qualitative agreement with that of the full-order model, even for a large $\Delta Re$.
Furthermore, the reconstruction error of the flow field estimated using the parametric ROM with pseudo-POD modes
decreased by decreasing the interval for subspace sampling in both the Reynolds-number and rotation-rate directions.
The distribution of the reconstruction error in the parameter space exhibited a trend similar to that of the subspace-estimation error,
which was affected by the subspace sensitivity.
This result indicates that obtaining an appropriate subspace for the target parameter was essential to improve the performance of the parametric ROM.
Specifically, our results implied that finely sampling subspaces in regions with high subspace sensitivity and coarsely sampling in regions with low sensitivity
was preferable for efficiently minimizing the reconstruction error with a limited number of subspace samples.

These findings were obtained from a geometric analysis of the POD modes on the matrix manifolds with the distance that relates the subspaces for different flow parameters.
When discussing the features of fluid-flow dynamics over a wide range of flow parameters (e.g., the initial and boundary conditions of the Navier--Stokes equations),
instead of attempting to extract a linear subspace in which the flow fields for the entire parameter space are described,
we showed that the subspaces indicating the local features of the fluid flow evolved continuously with changes in the parameters, thereby
providing another intuition for understanding fluid dynamics from a global perspective.
This perspective provided a framework for analyzing experimental and numerical simulation data using the geometric properties of the Grassmann manifold and Stiefel manifold,
which were interpreted abstractly.
Mode sensitivity analysis and parametric ROM will contribute to the extraction of comprehensive
insights into fluid dynamics over a wide range of flow parameters from data.
It will also lead to the development of novel techniques for the optimal design of fluid machinery, sensitivity analysis of fluid flows with respect to changes in parameters,
and active flow control, in which the evaluation of the flow field in a wide range of parameter spaces with significantly low computational costs is crucial.

\backsection[Funding]{This work was supported by JST PRESTO (Grant Number JPMJPR21O4) and JSPS KAKENHI (Grant Number 24K17442).}

\backsection[Declaration of interests]{The authors report no conflict of interest.}

\backsection[Author ORCIDs]{Shintaro Sato, https://orcid.org/0000-0002-9979-0051; Oliver T. Schmidt, https://orcid.org/0000-0002-7097-0235.}

\appendix
\section{Validity of rotation approximation of semi-orthogonal matrices}\label{Appendix1}

Here, we discuss the validity of the approximation in (\ref{eq:assumption}), which is used to define the sensitivity modes.
To evaluate this validity, we demonstrate that the semi-orthogonal matrix consisting of right singular vectors at a specified reference parameter $V^\mathrm{ref}$ 
can be represented as the product of the semi-orthogonal matrix $V(\xi)$ and $R^{T}(\xi)$.
The orthogonal matrix $R(\xi)$ is obtained by solving the orthogonal Procrustes problem \citep{Golub} as follows:
\begin{equation}
  \mathrm{minimize}~\|V(\xi)-V^\mathrm{ref}R(\xi)  \|_\mathrm{F},~~~\mathrm{subject~to}~R(\xi)\in O(r),\label{eq:Procrustes}
\end{equation}  
where $\|\cdot\|_\mathrm{F}$ indicates the Frobenius norm.
The SVD of $(V^\mathrm{ref})^TV(\xi)$ is performed to minimize $R(\xi)$ (\ref{eq:Procrustes}):
\begin{equation}
  (V^\mathrm{ref})^TV(\xi) \stackrel{\rm{SVD}}{=} U_\mathrm{P}\Sigma_\mathrm{P}V_\mathrm{P}^T.
\end{equation}
The product of $U_\mathrm{P}$ and $V_\mathrm{P}^T$ is the $R(\xi)$ that minimizes (\ref{eq:Procrustes}).

\begin{figure}
  \begin{center}
    \includegraphics[width=1.0\textwidth]{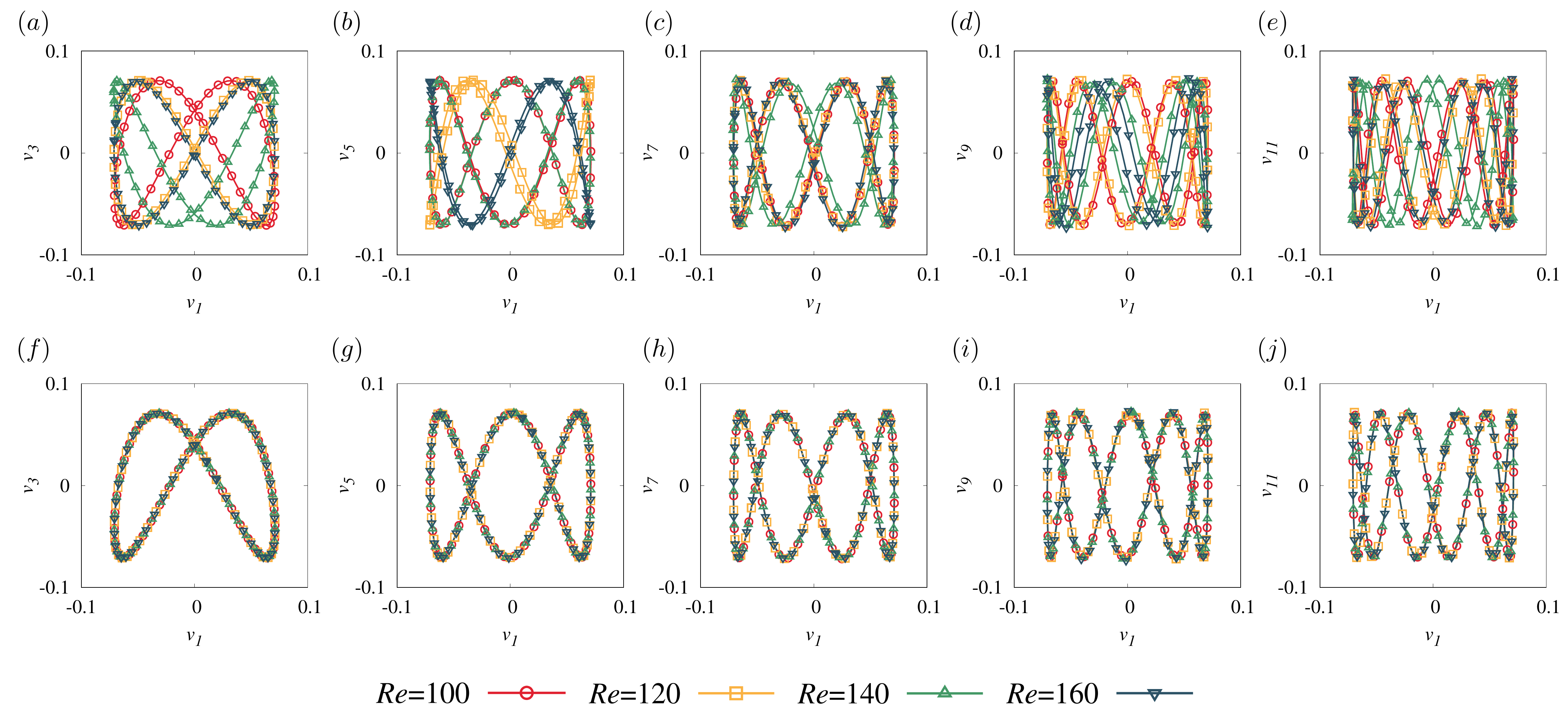}
    \caption{Phase portraits of the trajectories of expansion coefficients normalized by corresponding singular values for different Reynolds numbers at $\alpha=0.0$:
      (a,b,c,d,e) trajectories determined by $V(Re)$ (the parameter corresponds to the Reynolds number);
      (f,g,h,i,j) trajectories determined by $V(Re)R^T(Re)$.
      Panels (a,f), (b,g), (c,h), (d,i), and (e,j) show the phase portraits of the normalized expansion coefficients of the 1st--3rd, 1st--5th, 1st--7th, 1st--9th, and 1st--11th modes, respectively.
      \label{fig:A1}}
  \end{center}
\end{figure}
Figure~\ref{fig:A1}(a--e) show the phase portraits of the trajectories of $\boldsymbol{v}(t;\xi)\in\mathbb{R}^r$
for different Reynolds numbers ranging from $100$ to $160$ when the rotation rate was fixed at $0.0$,
where $V^T(\xi)=:\begin{bmatrix}\boldsymbol{v}_1(\xi) & \ldots & \boldsymbol{v}_{N_t}(\xi) \end{bmatrix}\in\mathbb{R}^{r\times N_t}$.
The product of $\boldsymbol{v}(t;\xi)$ and the singular values correspond to the expansion coefficients $\boldsymbol{a}(t;\xi)\in\mathbb{R}^r$.
The reference parameter is set to $(Re,\alpha)=(100,0.0)$.
The trajectory varies with the Reynolds number, whereas all trajectories exhibit periodic behavior (i.e. limit cycle).
This indicates that the difference in these periodic trajectories is owing to differences in their phases.
The phase portraits of the trajectories determined by the matrix $V(Re)R^T(Re)$ instead of $V(Re)$ are shown in figure~\ref{fig:A1}(f--j).
These trajectories coincide with that determined by $V^\mathrm{ref}$ regardless of the Reynolds number,
which suggests that the trajectories of $\boldsymbol{v}$ can be represented as $V(Re)\approx V^\mathrm{ref}R(Re)$.

\begin{figure}
  \begin{center}
    \includegraphics[width=1.0\textwidth]{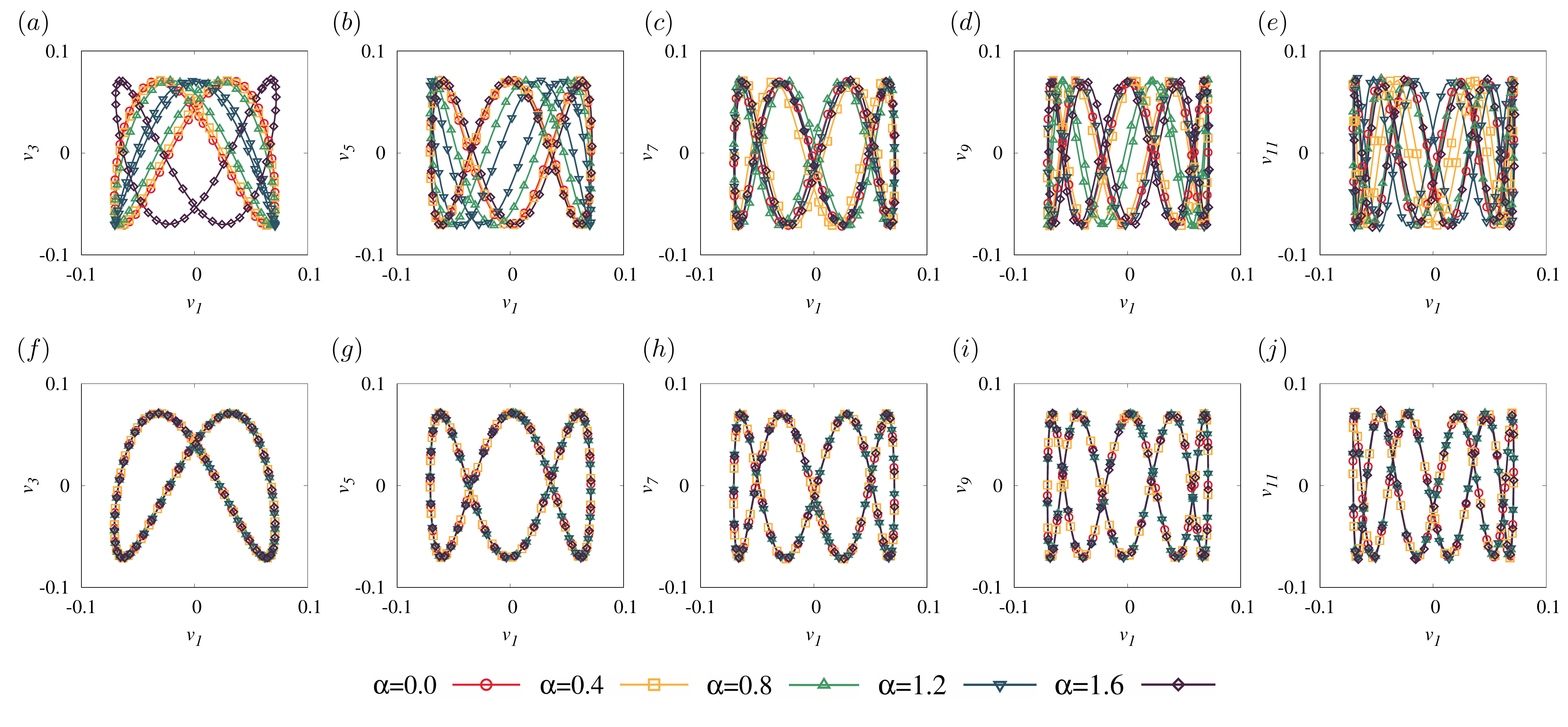}
    \caption{Phase portraits of the trajectories of the normalized expansion coefficients for different rotation rates at $Re=100$:
      (a,b,c,d,e) trajectories determined by $V(\alpha)$;
      (f,g,h,i,j) trajectories determined by $V(\alpha)R^T(\alpha)$.
      Panels (a,f), (b,g), (c,h), (d,i), and (e,j) show the phase portraits of the normalized expansion coefficients of the 1st--3rd, 1st--5th, 1st--7th, 1st--9th, and 1st--11th modes, respectively.\label{fig:A2}}
  \end{center}
\end{figure}
Figure~\ref{fig:A2}(a--e) show phase portraits of the trajectory for rotation rates ranging from $0.0$ to $1.6$
when the Reynolds number is fixed at $100$.
The trajectory, which shows a limited cycle, varies with the rotation rate, as in the case of the Reynolds-number variation.
The trajectories determined by the matrix $V(\alpha)R^T(\alpha)$ show good agreement with the trajectory of $V^\mathrm{ref}$,
where the reference parameter is $(Re,\alpha)=(100,0.0)$, as shown in figure~\ref{fig:A2}(f--j).
Thus, we can conclude that matrix $R(\xi)$ can be interpreted as representing the phase differences between the trajectories determined by $V(\xi)$ and $V^\mathrm{ref}$,
and the matrix $V(\xi)$ can be approximated as the product of $V^\mathrm{ref}$ and $R(\xi)$.
This approximation is valid for the flow parameters considered in this study
because the difference in trajectories is caused by the phase difference in these trajectories.

\section{Error evaluations of subspace estimation and flow field reconstruction: one-parameter variation case}\label{Appendix2}

Here, we evaluate the errors of the subspace estimation and flow field reconstruction over a wide range of Reynolds numbers
using a parametric ROM based on subspace interpolation on the Grassmann manifold. 
Parametric ROMs constructed using subspaces sampled at different $\Delta Re$ values are examined to discuss convergence of the estimation error
with respect to the subspace-sampling interval $\Delta Re$.
\begin{table}
  \begin{center}
    {\tabcolsep = 0.3cm
    \begin{tabular}{ccc}
      $\Delta Re$ & $Re$ for training dataset & $Re$ for test dataset \\
      10 & 60, 70, 80, 90, 100, 110, 120, 130, 140 & 65, 75, 85, 95,105, 115, 125, 135\\
      20 & 60, 80, 100, 120, 140 & 70, 90, 110, 130\\
      30 & 60, 90, 120, 150 & 75, 105, 135\\
      40 & 60, 100, 140 & 80, 120 \\
    \end{tabular}
    }
    \caption{Sampling points $Re$ used as the training dataset for POD modes and Reynolds numbers for evaluation (denoted as the test dataset) for each $\Delta Re$.}
    \label{tab:sampling}
  \end{center}
\end{table}
Table~\ref{tab:sampling} summarizes the conditions for $\Delta Re$ considered in this study, the Reynolds numbers used to interpolate the subspaces (denoted as the training dataset),
and the Reynolds numbers used to estimate the subspaces and reconstruct the flow field (denoted as the test dataset).
The Reynolds numbers for subspace and flow field estimation are determined so that their average coincides with the Reynolds number of the training dataset,
i.e., the Reynolds number for the estimation is $Re=(Re_1+Re_2)/2$ when using the subspaces at $Re_1$ and $Re_2(=Re_1+\Delta Re)$.

\begin{figure}
  \begin{center}
    \includegraphics[width=1.0\textwidth]{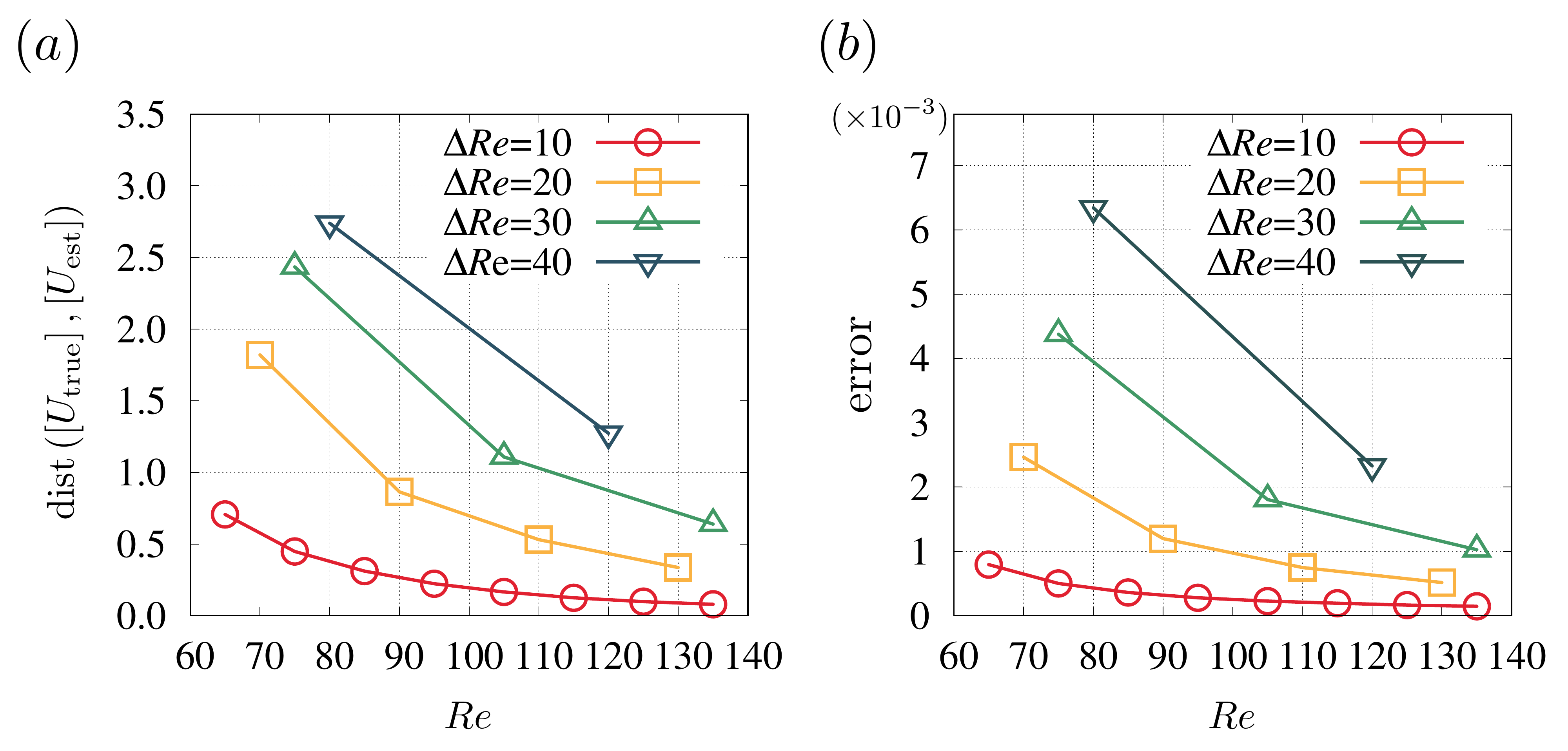}
    \caption{Error evaluation of the parametric ROM for a wide range of the Reynolds numbers using different $\Delta Re$;
      (a) subspace error; (b) flow field reconstruction error.\label{fig:fig15}}
  \end{center}
\end{figure}
Figure~\ref{fig:fig15}(a) shows the subspace-estimation errors as functions of the Reynolds number for different $\Delta Re$ values.
The subspace-estimation error decreases with decreasing $\Delta Re$ for all Reynolds numbers considered in this study.
In addition, the error decreases as increasing the Reynolds number, regardless of $\Delta Re$,
because the subspace sensitivity decreases as increasing the Reynolds number (see figure~\ref{fig:SubspaceSens}). 
The error in the flow field reconstructed using the parametric ROM as a function of the Reynolds number for different $\Delta Re$ values is shown in figure~\ref{fig:fig15}(b).
The reconstruction error decreases with decreasing $\Delta Re$ and increasing Reynolds number for a fixed $\Delta Re$.
This highlights that the reconstruction error is closely related to the subspace-estimation error.


\bibliographystyle{jfm}

\begin{thebibliography}{99}

\expandafter\ifx\csname natexlab\endcsname\relax
\def\natexlab#1{#1}\fi
\expandafter\ifx\csname selectlanguage\endcsname\relax
\def\selectlanguage#1{\relax}\fi
\bibitem[Absil {\it et al.} (2008)]{Absil}
  {\sc Absil, P. A., Mahony, R. \& Sepulchre, R.} 2008
  {\it Optimization algorithms on matrix manifolds}.
  Princeton University Press.
\bibitem[Ahlborn {\it et al.} (2002)]{Ahlborn2002}
  {\sc Ahlborn, B., Seto, M. L. \& Noack, B. R.} 2002
  {On drag, Strouhal number and vortex-street structure}.
  {\it Fluid Dyn. Res.} {\bf 30}, 379--399.
\bibitem[Amsallem \& Farhat (2008)]{Amsallem2008}
  {\sc Amsallem, D. \& Farhat, C.} 2008
  {Interpolation method for adapting reduced-order models and application to aeroelasticity}.
  {\it AIAA J.} {\bf 46} (7), 1803--1813.
\bibitem[Bendokat {\it et al.} (2024)]{Bendokat2024}
  {\sc Bendokat, T., Zimmermann, R. \& Absil, P. A.} 2024
  {A Grassmann manifold handbook: basic geometry and computational aspects}.
  {\it Adv. Compt. Math.} {\bf 50}, 6
\bibitem[Benner {\it et al.} (2015)]{Benner2015}
  {\sc Benner, P., Gugercin, S. \& Willcox, K.} 2015
  {A survey of projection-based model reduction methods for parametric dynamical systems}.
  {\it SIAM Rev.} {\bf 57} (4), 483--531.
\bibitem[Boumal \& Absil (2015)]{Boumal2015}
  {\sc Boumal, N. \& Absil, P. A.} 2015
  {Low-rank matrix completion via preconditioned optimization on the Grassmann manifold}.
  {\it Linear Algebra Appl.} {\bf 475}, 200--239.
\bibitem[Edelman {\it et al.} (1998)]{Edelman1998}
  {\sc Edelman, A., Arias, T. A. \& Smith, S. T.} 1998
  {The geometry of algorithms with orthogonality constraints}.
  {\it SIAM J. Matrix Aanl. Appl.} {\bf 20} (2), 303--353.
\bibitem[Galletti {\it et al.} (2004)]{Galletti2004}
  {\sc Galletti, B., Bruneau, C. H., Zannetti, L. \& Iollo, A.} 2004
  {Low-order modelling of laminar flow regimes past a confined square cylinder}.
  {\it J. Fluid Mech.} {\bf 503}, 161--170.
\bibitem[Gulub \& Loan (2013)]{Golub}
  {\sc Golub, G. H. \&  van Loan, C. F.} 2013
  {\it Marix Computations}. 4th ed.
  Johns Hopkins University Press.
\bibitem[Hay {\it et al.} (2009)]{Hay2009}
  {\sc Hay, A., Borggaard, J. T. \& Pelletier, D.} 2009
  {Local improvements to reduced-order models using sensitivity analysis of the proper orthogonal decomposition}.
  {\it J. Fluid Mech.} {\bf 629}, 41--72.
\bibitem[Hess {\it et al.} (2023)]{Hess2023}
  {\sc Hess, M. W., Quaini, A. \& Rozza, G.} 2023
  {A data-driven surrogate modeling approach for time-dependent incompressible Navier-Stokes equations with dynamic mode decomposition and manifold interpolation}
  {\it Adv. Comput. Math.} {\bf 49}, 22.
\bibitem[H\"{u}per {\it et al.} (2021)]{Huper2021}
  {\sc H\"{u}per, K., Markina, I. \& Leite, F.S.} 2021
  {A Lagrangian approach to external curves on Stiefel manifolds}.
  {\it J. Geom. Mech.} {\bf 13} (1), 55--72.
\bibitem[Lee (2018)]{Lee}
  {\sc Lee, J. M.} 2018
  {\it Introduction to Riemannian manifolds}.
  Graduate Texts in Mathematics, vol. 176. Springer. 
\bibitem[Lele (1992)]{Lele1992}
  {\sc Lele, S. K.} 1992
  {Compact finite difference schemes with spectral-like resolution}.
  {\it J. Compt. Phys.} {\bf 103} (1), 16--42.
\bibitem[Lieu \& Farhat (2007)]{Lieu2007}
  {\sc Lieu, T. \& Farhat, C.} 2007
  {Adaptation of aeroelastic reduced-order models and application to an F-16 configuration}.
  {\it AIAA J.} {\bf 45} (6), 1244--1257.
\bibitem[Liu \& Liu (2022)]{Liu2022}
  {\sc Liu, X. \& Liu, X.} 2022
  {Regression trees on Grassmann manifold for adapting reduced-order models}.
  {\it AIAA J.} {\bf 61} (3), 1318--1333.
\bibitem[Lui (2012)]{Lui2012}
  {\sc Lui, Y. M.} 2012
  {Advances in matrix manifolds for computer vision}
  {\it Image Vis. Comput.} {\bf 30}, 380--388.
\bibitem[Lumley (1970)]{Lumley}
  {\sc Lumley, J. L.} 1970
  {\it Stochastic Tools in Turbulence}.
  Academic Press.
\bibitem[Ma \& Karniadakis (2002)]{Ma2002}
  {\sc Ma, X. \& Karniadakis, G. E.} 2002
  {A low-dimensional model for simulating three-dimensional cylinder flow}.
  {\it J. Fluid Mech.} {\bf 458}, 181--190.
\bibitem[Nakamura {\it et al.} (2024)]{Nakamura2024}
  {\sc Nakamura, Y., Sato, S. \& Ohnishi, N.} 2024
  {Application of proper orthogonal decomposition to flow fields around various geometries and reduced-order modeling}.
  {\it Comput. Methods Appl. Mech. Engrg.} {\bf 432} (1), 117340.
\bibitem[Noack \& Eckelmann (1994)]{Noack1994}
  {\sc Noack, B. R. \& Eckelmann, H.} 1994
  {A global stability analysis of the steady and periodic cylinder wake}.
  {\it J. Fluid Mech.} {\bf 270}, 297--330.
\bibitem[Noack {\it et al.} (2003)]{Noack2003}
  {\sc Noack, B. R., Afansasievm K., Morzynski, M., Tadmor, G \& Thieke, F.}
  {A hierarchy of low-dimensional models for the transient and post-transient cylinder wake}.
  {\it J. Fluid Mech.} {\bf 497}, 335--363.
\bibitem[Noack {\it et al.} (2011)]{Noack2011}
  {\sc Noack, B. R., Morzy\'nski, M. \& Tadmor, G.} 2011
  {\it Reduced-Order Modelling for Flow Control}.
  vol. 528. Springer Science \& Business Media.
\bibitem[Pawar {\it et al.} (2020)]{Pawar2020}
  {\sc Pawar, S., Ahmed, S. E., San, O. \& Rasheed, A.} 2020
  {Data-driven recovery of hidden physics in reduced order modeling of fluid flows}.
  {\it Phys. Fluids} {\bf 32}, 036602.
\bibitem[Pelletier {\it et al.} (2008)]{Pelletier2008}
  {\sc Pelletier, D., Hay, A., Etienne, S. \& Borggaard, J.} 2008
  {The sensitivity equation method in fluid mechanics}.
  {\it Eur. J. Compt. Mech.} {\bf 17} (1--2), 31--61.
\bibitem[Rowley \& Dawson (2017)]{Rowley2017}
  {\sc Rowley, C. W. \& Dawson, S. T. M.} 2017
  {Model reduction for flow analysis and control}.
  {\it Annu. Rev. Fluid Mech.} {\bf 49}, 387--417.
\bibitem[Sato {\it et al.} (2021)]{Sato2021}
  {\sc Sato, S., Sakamoto, H. \& Ohnishi, N.} 2021
  {Connections between the modes of a nonlinear dynamical system on a manifold}.
  {\it Phys. Rev. E} {\bf 103}, 062210.
\bibitem[Schmid (2010)]{PSchmid2010}
  {\sc Schmid, P. J.} 2010
  {Dynamic mode decomposition of numerical and experimental data}.
  {\it J. Fluid Mech.} {\bf 656}, 5--28.
\bibitem[Sierra {\it et al.} (2020)]{Sierra2020}
  {\sc Sierra, J., Fabre, D., Citro, V. \& Giannetti, F.} 2020
  {Bifurcation scenario in the two-dimensional laminar flow past a rotating cylinder}.
  {\it J. Fluid Mech.} {\bf 905}, A2.
\bibitem[Sirovich (1987)]{Sirovich1987}
  {\sc Sirovich, L} 1987
  {Turbulence and the dynamics of coherent structures I. Coherent structures}.
  {\it Q. Appl. Math.} {\bf 45} (3), 561--571.
\bibitem[Son (2013)]{Son2013}
  {\sc Son, N. T.} 2013
  {A real time procedure for affinely dependent parametric model order reduction using interpolation on Grassmann manifolds}.
  {\it Int. J. Numer. Meth. Eng.} {\bf 93}, 818--833.
\bibitem[Taira {\it et al.} (2017)]{Taira2017}
  {\sc Taira, K., Brunton, S. L., Dawson, S. T. M., Rowley, C. W., Colonius, T., McKeon, B. J., Schmidt, O. T., Gordeyev, S., Theofilis, V. \& Ukeiley, L. S.} 2017
  {Modal analysis of fluid flows: An overview}.
  {\it AIAA J.} {\bf 55} (12), 4013--4041.
\bibitem[Taylor \& Glauser (2004)]{Taylor2004}
  {\sc Taylor, J. A. \& Glauser, M. N.} 2004
  {Towards practical flow sensing and control via POD and LSE based low-dimensional tools}.
  {\it Trans. ASME J. Fluids Engng.} {\bf 126} (3), 337--345.
\bibitem[Towne {\it et al.} (2018)]{Towne2018}
  {\sc Towne, A., Schmidt, O. T. \& Colonius, T.} 2018
  {Spectral proper orthogonal decomposition and its relationship to dynamic mode decomposition and resolvent analysis}
  {\it J. Fluid Mech.} {\bf 847}, 821--867.
\bibitem[Visbal \& Gaitonde (2002)]{Visbal2002}
  {\sc Visbal, M.R. \& Gaitonde, D.V.} 2002
  {On the use of higher-order finite-difference schemes on curvilinear and deforming meshes}.
  {\it J. Compt. Phys.} {\bf 181} (1), 155--185.
\bibitem[Williamson \& Brown (1998)]{Williamson1998}
  {\sc Williamson, C.H.K. \& Brown, G.L.} 1998
  {A series in $1/\sqrt{Re}$ to represent the Strouhal-Reynolds number relationship to the cylinder wake}.
  {\it J. Fluids Struct.} {\bf 12} (8), 1073--1085.
\bibitem[Wong (1967)]{Wong1967}
  {\sc Wong, Y.C.} 1967
  {Differential geometry of Grassmann manifolds}.
  {\it Proc. Natl. Acad. Soc.} {\bf 57} (3), 589--594.
\bibitem[Yoon \& Jameson (1988)]{Yoon1988}
  {\sc Yoon, S. \& Jameson, A.} 1988 {Lower-upper symmetric-Gauss-Seidel method for the Euler and Navier-Stokes equations}.
  {\it AIAA J.} {\bf 26} (9), 1025--1026.
\bibitem[Zimmermann (2017)]{Zimmermann2017}
  {\sc Zimmermann, R.} 2017
  {A matrix-algebraic algorithm for the Riemannian logarithm on the Stiefel manifold under the canonical metric}.
  {\it SIAM J. Matrix Anal. Appl.} {\bf 38} (2), 322--342.
\bibitem[Zimmermann (2019)]{Zimmermann2019}
  {\sc Zimmermann, R.} 2019
  {Manifold interpolation and model reduction}. {\it arXiv preprint}, 1902.06502.
\end{thebibliography}

\end{document}